\documentclass[aip,jcp,amsmath,amssymb,reprint]{revtex4-1}

\usepackage{graphicx}
\usepackage{graphicx,color}
\usepackage{dcolumn}
\usepackage{bm}

\begin{document}

\preprint{AIP/123-QED}

\title{Phase behaviour of semiflexible lattice polymers in poor-solvent solution: mean-field theory and Monte Carlo simulations}

\author{Davide Marcato}
\email{dmarcato@sissa.it}
\affiliation{Scuola Internazionale Superiore di Studi Avanzati (SISSA), Via Bonomea 265, 34136 Trieste, Italy}
\author{Achille Giacometti}
\email{achille.giacometti@unive.it}
\affiliation{Dipartimento di Scienze Molecolari e Nanosistemi, Universit\`a Ca' Foscari Venezia, 30123 Venezia, Italy}
\affiliation{European Centre for Living Technology (ECLT) Ca' Bottacin, 3911 Dorsoduro Calle Crosera, 30123 Venezia, Italy}
\author{Amos Maritan}
\email{amos.maritan@pd.infn.it}
\affiliation{Laboratory of Interdisciplinary Physics, Department of Physics and Astronomy ``G. Galilei'', University of Padova, Padova, Italy and INFN, Sezione di Padova, via Marzolo 8, 35131 Padova, Italy}
\author{Angelo Rosa}
\email{anrosa@sissa.it}
\affiliation{Scuola Internazionale Superiore di Studi Avanzati (SISSA), Via Bonomea 265, 34136 Trieste, Italy}


\date{\today}

\begin{abstract}
We study a solution of interacting semiflexible polymers with curvature energy in poor-solvent conditions on the $d$-dimensional cubic lattice using mean-field theory and Monte Carlo computer simulations.
Building upon past studies on a single chain, we construct a field-theory representation of the system and solve it within a mean-field approximation supported by Monte Carlo simulations in $d=3$.
A gas-liquid transition is found in the temperature-density plane that is then interpreted in terms of real systems.
Interestingly, we find this transition to be independent of the bending rigidity.
Past classical Flory-Huggins and Flory mean-field results are shown to be particular cases of this more general framework.
Perspectives in terms of guiding experimental results towards optimal conditions are also proposed.
\end{abstract}

\maketitle

\section{Introduction}\label{sec:intro}
Determining the phase behaviour of a solution of flexible and semiflexible polymers in poor-solvent conditions is a particularly challenging problem for several reasons.
Unlike the case of colloidal liquids, where unambiguous gas-liquid and liquid-solid transitions are theoretically well  characterized~\cite{Hansen2006} and experimentally observed~\cite{Pusey1986}, in the case of polymer solutions the presence of chain connectivity~\cite{Flory1953,deGennes1979,DoiEdwards1988,Rubinstein2003} makes a full understanding of its phase behavior much more challenging, in particular at high concentrations.

One of the emerging conceptual problems hinges on the difficulty to discriminate between purely kinetic effects and those associated with the underlying thermodynamics~\cite{Kawak2021,Wilken-PRX2023}.
For instance, polymers do not completely crystallize when cooled down but become structured into a hierarchy of ordered structures~\cite{Olmsted1998}.
Also, it has been argued that on cooling a polymer melt undergoes a spinodal decomposition thus making the crystallization metastable and leaving the system out of equilibrium~\cite{Kawak2021}.
Another difficulty stems from the large number of thermodynamic and structural parameters that need to be taken into consideration:
in fact, in addition to usual thermodynamic quantities such as temperature and pressure that control the system, many other microscopic parameters such as {\it interchain} (in addition to the {\it intrachain}) interactions, the number of monomers in a chain, the stiffness of the fiber and the total polymer volume fraction have to be taken into account~\cite{Flory1953,deGennes1979,DoiEdwards1988,Rubinstein2003} and become axes of a large parameter space.

Particularly important appears to be the case of semiflexible polymers, as a paradigmatic example of the protein folding problem~\cite{Bascle1992,Bascle1993a,Bascle1993b,Doniach1996} or biopolymers in a crowded environment~\cite{Ranganathan2022}, and this is currently stimulating many studies along these lines.
Our current understanding of these system comes, in particular, from computer simulations which, however, have been limited so far to particular situations. 
A first group of studies derived the gas-liquid phase diagram for flexible~\cite{Sheng1994} and semiflexible~\cite{Sheng1996} bead-spring chains up to only $100$ monomers per chain using Monte Carlo (MC) simulations.
In particular, they found a phase-equilibrium diagram very similar to that of simple liquids with some minor effects ascribed to the bending rigidity.
Similar results were obtained by more recent and extensive simulations~\cite{Ivanov2003}.
Other simulations aimed at understanding entanglement properties between distinct chains~\cite{Bobbili2020,Everaers2020,Dietz2022} or the onset of nucleation process~\cite{Kos2021}.
However, a comprehensive picture of the phase behavior of semiflexible polymers in poor-solvent solutions is currently lacking.

Surprisingly, even in the case of a single semiflexible polymer, a general theoretical understanding of the phase behaviour is still lacking notwithstanding several studies with different techniques have recently appeared~\cite{Montesi2004,Seaton2013,Chertovich2014,Marenz2016,Werlich2017,Skrbic2019} that remained, however, focused on rather specific questions. 
For instance, by using mean-field arguments supported by bead-spring MC simulations, it has been argued that the ground state of a single semiflexible chain can be either a rod-like or a toroidal structure depending on the bending rigidity and the contour length of the polymer~\cite{Hoang2014,Hoang2015} and this has been confirmed recently~\cite{Aierken2023} by computer simulations which also accounts for the temperature dependence.
Interestingly, classical studies of a lattice model~\cite{Bastolla1997,Doye1998} observe only rod-like phases, in the form of Hamiltonian paths, likely due to the geometrical constraints imposed by the lattice.
Remarkably, the observed phase diagram is in excellent agreement with mean-field predictions using a field theoretical approach~\cite{Doniach1996}.

While a full understanding of the differences observed in lattice and off-lattice simulations is an interesting issue on its own right (and that will be discussed elsewhere), the present study will focus on providing the multi-chains counterpart of the aforementioned single chain studies~\cite{Doniach1996,Bastolla1997,Doye1998}.
Specifically, in the wake of the long-standing tradition of lattice models for modeling polymer structure~\cite{Flory1956a,deGennes1972,desCloizeaux1975,Gujrati1980a,GujratiMagneticAnalog1981,Gujrati1981a,WheelerPfeuty1981,Gujrati1982a,FreedMultiChain1985,Orland1985,Bascle1992,Bascle1993a,Bascle1993b,Doniach1996,JacobsenKondev1998,HiguchiPRE1998,ZilmanSafranPRE2022}, here we develop a field-theoretical description of semiflexible self-avoiding chains with attractive interactions on the $d$-dimensional hypercubic lattice, and solve it within a mean-field approximation.
Lattice grand canonical MC simulations will then be presented demonstrating the remarkable accuracy of the mean-field predictions.
The multi-chain field theory approach that is proposed here build on past work by des Cloizeaux~\cite{desCloizeaux1975} that extends the classical relation pointed out long ago by de Gennes~\cite{deGennes1972,deGennes1979} of self-avoiding walks as the $n \to 0$ limit of a spin $O(n)$-model, $n$ being the number of components of each spin on a lattice point.

Notwithstanding its limitation, the present lattice approach has the great merit of making transparent the underlying physics and provide a guidance of the regions of this large parameter space that warrant a more in-depth analysis with dedicated numerical or experimental techniques.
As a by-product of the theoretical analysis, we will re-obtain some classical results within a wider framework, as we will see.

The paper is organized as the following.
In Section~\ref{sec:StateOfTheArt},
we provide a concise summary of the current understanding of the phase behavior of interacting semiflexible polymers.
In particular, due to its specialized nature and for having inspired this work, we are going to highlight some salient aspects related to the past work of Orland and colleagues~\cite{Orland1985,Bascle1992,Doniach1996}.
Moreover, we will also discuss some conclusions from recent computational work by other groups which has the advantage of providing a broader view beyond the mean-field approach.
The novel part of this work starts in Sec.~\ref{sec:LatticeModel-Definition},
where we will introduce the exact grand canonical partition function $Z$ of the lattice model for a multi-chain system ({\it i.e.}, a polymer solution) on the $d$-dimensional cubic lattice which takes into account the local bending stiffness of the polymer fiber, excluded-volume and short-range attractive interactions between close-by monomers.
Then, by exploiting the analogy between self-avoiding polymers and the spin $O(n \! \rightarrow \! 0)$-model, we construct (Sec.~\ref{sec:LatticeModel-FieldTheory})
the exact field-theoretic representation of $Z$.
Since an exact computation of $Z$ is unfeasible, we describe a uniform saddle-point approximation (Sec.~\ref{sec:LatticeModel-MFApprox})
and obtain the corresponding mean-field solution of the problem, the reliability of which is demonstrated by comparison to MC computer simulations (Secs.~\ref{sec:LatticeModel-MC} and~\ref{sec:MFvsMC-Results}).
Finally we show that our results (Sec.~\ref{sec:FH-equivalence})
recapitulate, as particular cases, several models that have been discussed in the past and we demonstrate the equivalence between our approach and the classical Flory-Huggins theory~\cite{Flory1942,Huggins1942} for mixtures.
Discussion and conclusions, with an outline on open problems and possible future perspective, are presented in Sec.~\ref{sec:Concls}.

\section{Review of the single-chain formalism: Hamiltonian rings}\label{sec:StateOfTheArt}
Before introducing (Sec.~\ref{sec:LatticeModel})
our field-theoretic formalism for semiflexible polymer solutions and in order to set the stage, it proves instructive to recapitulate the single chain formalism~\cite{Orland1985,Bascle1992,Doniach1996}.

Denoting by $L$ the linear size of the hypercubic lattice in $d$ dimensions and by $a$ the lattice spacing, 
we consider {\it Hamiltonian paths}, polymer chains whose number of nodes $N$ is equal to $V/a^d = (L/a)^d$ where $V$ is the volume of the lattice: that is, we consider all lattice points occupied and no vacancies.
For simplicity, we further restrict our considerations to {\it closed paths}, or {\it Hamiltonian rings} (HR), knowing that the statistics is the same in the thermodynamic limit~\cite{Orland1985}.

Consider the following ${\mathcal O}(n)$-vector model where the $n$-component vector,
\begin{equation}\label{eq:SpinVectors}
{\vec S}({\vec x}) \equiv \left( S^1\!({\vec x}), S^2\!({\vec x}), ..., S^n\!({\vec x}) \right) \, ,
\end{equation}
is associated to each lattice point ${\vec x}$.
By defining the scalar product
${\vec S}(\vec x) \cdot {\vec S}(\vec x^{\prime}) \equiv \sum_{i=1}^n S^i\!(\vec x) S^i\!(\vec x^{\prime})$
between any two vectors associated to lattice points $\vec x$ and $\vec x^{\prime}$, we assume the following constraint on the norm-square of $\vec S(\vec x)$:
\begin{equation}\label{eq:n-constraint}
{\vec S}(\vec x)^2 \equiv {\vec S}(\vec x) \cdot {\vec S}(\vec x) =  \sum_{i=1}^n S^i\!(\vec x)^2 = n \, .
\end{equation}
The reduced Hamiltonian for the $n$-vector model in zero external field reads
\begin{equation}\label{eq:hamiltonian}
-\beta {\mathcal H} = \frac{J}2 \, \sum_{\vec x, \vec x^{\prime}} \Delta(\vec x, \vec x^{\prime}) \, {\vec S}(\vec x) \cdot {\vec S}(\vec x^{\prime}) \, , 
\end{equation}
where
$\beta=1/(k_B T)$, $k_B$ being the Boltzmann constant and $T$ being the temperature,
$J$ is the coupling constant between the spins,
and where
\begin{equation}\label{eq:def_delta}
\Delta(\vec x, \vec x^{\prime}) =
\left\{
\begin{array}{cl}
1 \, , & \text{if} \, \, \, |\vec x - \vec x^{\prime}| = a \\
0 \, , & \text{otherwise} 
\end{array}
\right. \, .
\end{equation}
From now on we assume periodic boundary conditions and Eq.~\eqref{eq:hamiltonian} can be also written as
\begin{equation}\label{eq:hamiltonian-pbc}
-\beta {\mathcal H} = J \, \sum_{\vec x} \sum_{\mu=1}^d {\vec S}(\vec x) \cdot {\vec S}({\vec x} + {\hat e}_{\mu}) \, , 
\end{equation}
where ${\hat e_{\mu}}$ is the unit lattice vector pointing towards the ``$+\mu$''-direction.
Then, we introduce the integration measure,
\begin{equation}\label{eq:integration_measure}
d\Omega_n(\vec x) = d{\vec S}(\vec x) \, \delta\!\left( \sum_{i=1}^n S^i\!(\vec x)^2 - n \right) \, ,
\end{equation}
with the Dirac $\delta$-function enforcing the constraint Eq.~\eqref{eq:n-constraint} and 
the related {\it geometrical} average (or, {\it trace} operation)
\begin{equation}\label{eq:geometrical_average}
\langle ... \rangle_{\Omega} \equiv \frac{\int \prod_{\vec x} d\Omega_n(\vec x) \, (...)}{\int \prod_{\vec x} d\Omega_n(\vec x)} \, .
\end{equation}
A very peculiar feature of the operation defined in Eq.~\eqref{eq:geometrical_average} is that, in the formal limit $n \to 0$, the following equality for the moment-generating function holds~\cite{deGennes1979}:
\begin{equation}\label{eq:moment_theorem}
\left\langle \prod_{\vec x} \, e^{{\vec S}({\vec x}) \cdot {\vec \varphi}(\vec x)} \right\rangle_{\Omega} = \prod_{\vec x} \left(1 + \frac12 \, {\vec \varphi}(\vec x)^2 \right) \, ,
\end{equation}
{\it i.e.}, the moment-generating function has a simple quadratic form.
Further consequences of this fact will be illustrated in Appendix~\ref{App:Exp-BetaH-expansion}.

Let us consider now the quantity $\langle e^{-\beta H}\rangle_{\Omega}$.
One can show that, in the formal limit $n\rightarrow0$, the following expansion holds:
\begin{equation}\label{eq:geometrical_partition_function}
\left \langle e^{-\beta {\mathcal H}} \right \rangle_{\Omega} = 1 + n\left(\sum_{\ell=1}^{N} J^\ell \, Z_\ell \! \right) + {\mathcal O}\!\left(n^2\right) \, ,
\end{equation}
where $Z_\ell$ is the total number of self-avoiding closed paths of total length $\ell$ (extended details on the derivation of Eq.~\eqref{eq:geometrical_partition_function}, which is non-trivial, are provided in Appendix~\ref{App:Exp-BetaH-expansion}). 
At the same time, by using the standard Hubbard-Stratonovich transformation~\cite{HubbardPRL1959,Chaikin00}, we obtain
\begin{equation}\label{eq:HS_transformation}
\left \langle e^{-\beta {\mathcal H}} \right \rangle_{\Omega} = \frac{\int \prod_{\vec x} d{\vec \varphi}(\vec x) \, e^{-A} \, \left\langle e^{\sqrt{J} \sum_{\vec x} {\vec S}(\vec x) \cdot {\vec \varphi}(\vec x)} \right\rangle_{\Omega}}{\int \prod_{\vec x}
d{\vec \varphi}(\vec x) \, e^{-A}} \, ,
\end{equation}
where
\begin{equation}\label{eq:function_A}
A = \frac12 \sum_{\vec x, \vec x^{\prime}} \Delta^{-1} \! \left( \vec x, \vec x^{\prime} \right) \vec{\varphi}\!(\vec x) \cdot \vec{\varphi}\!(\vec x^{\prime}) \, ,
\end{equation}
with $\Delta^{-1}$ the inverse of the matrix $\Delta$ (Eq.~\eqref{eq:def_delta}).
At this point it has to be noted that, strictly speaking, the matrix $\Delta$ is not positive definite, therefore the Hubbard-Stratonovich transformation itself is, in principle, ill-defined.
However, this technical difficulty can be overcome through the more rigorous approach~\cite{Orland1985,HiguchiPRE1998} involving Fresnel integrals which leaves the final results unaffected.
In the end then and in the limit $n\rightarrow0$, the numerator of Eq.~\eqref{eq:HS_transformation} can be easily computed by resorting to Eq.~\eqref{eq:moment_theorem}, granting the result
\begin{equation}\label{eq:results}
\langle e^{-\beta {\mathcal H}} \rangle_{\Omega} = \frac{\int \prod_{\vec x} d{\vec \varphi}(\vec x) \, e^{-A} \, \prod_{\vec x} \left( 1+\frac{J}2 {\vec \varphi}(\vec x)^2 \right)}{\int \prod_{\vec x}
d{\vec \varphi}(\vec x) \, e^{-A}} \, .
\end{equation}
We now notice that $Z_N$ in Eq.~\eqref{eq:geometrical_partition_function} -- which coincides with the {\it total} number of HR on our lattice -- can be formally~\cite{FranzOrland1999} obtained as
\begin{equation}\label{eq:}
Z_N = \lim_{n\rightarrow0} \lim_{J\rightarrow\infty} \frac1n \frac1{J^N} \langle e^{-\beta {\mathcal H}} \rangle_{\Omega} \, .
\end{equation}
Finally, by combining Eq.~\eqref{eq:results} with~\eqref{eq:} we get the following expression for $Z_N$,
\begin{equation}\label{eq:final}
Z_N
= \lim_{n \to 0} \frac1n \frac{\int \prod_{\vec x} d{\vec \varphi}(\vec x) \, e^{-A} \, \prod_{\vec x} \frac12 {\vec \varphi}(\vec x)^2}{\int \prod_{\vec x} d{\vec \varphi}(\vec x) \, e^{-A}} \, ,
\end{equation}
which was introduced first in~\cite{Orland1985}.
It is important to stress that, in order to compute $Z_N$, we have used the fact that the trace operation Eq.~\eqref{eq:geometrical_average} has the very peculiar properties described in Appendix~\ref{App:Exp-BetaH-expansion}.
Finally, for the purpose of computing $Z_N$, the geometrical origin of this trace ({\it i.e.}, constraining the spin vectors on the surface of a sphere of radius $\sqrt{n}$) becomes completely irrelevant.

We demonstrate now that there exists an alternative method~\cite{LisePhD} of finding $Z_N$~\eqref{eq:final} that represents both a shortcut with respect to the approach presented so far and has the great advantage of being exportable to the more general situation considered here (Sec.~\ref{sec:LatticeModel})
in a relatively straightforward manner. 

The method consists in defining {\it a priori} a trace operation -- that we denote by the symbol $\langle \cdot \rangle_{0}$ -- characterized by the desired mathematical properties:
\begin{eqnarray}
\langle 1 \rangle_0 & = & 0 \, , \label{eq:IntroducingTraceProperties-1} \\
\langle S^i \rangle_0 & = & 0 \, , \label{eq:IntroducingTraceProperties-2} \\
\langle S^i S^j \rangle_0 & = & \delta_{ij} \, , \label{eq:IntroducingTraceProperties-3} \\
\langle S^{i_1} S^{i_2} \, ... \,  S^{i_k} \rangle_0 & = & 0 \, , \, \, \, \text{if $k \geq 3$} \, , \label{eq:IntroducingTraceProperties-4}
\end{eqnarray}
between spin components {\it on the same} lattice site, while $S$-vectors {\it on different} sites are independent from each other.
Notice that the only difference from the trace~\eqref{eq:geometrical_average} is that now the trace of $1$ is equal to $0$ (see Appendix~\ref{App:Exp-BetaH-expansion}).

Based on the definitions~\eqref{eq:IntroducingTraceProperties-1}-\eqref{eq:IntroducingTraceProperties-4} and by taking $J=1$ in the Hamiltonian of the $n$-vector model (Eq.~\eqref{eq:hamiltonian}), the partition function~\eqref{eq:final} of the HR is equivalent to:
\begin{equation}\label{eq:Lise-CanonicalZ-FieldFormulation}
Z_N = \lim_{n \rightarrow 0} \frac1n \left\langle e^{-\beta {\mathcal H}} \right\rangle_0 \, .
\end{equation}
In fact, by taking the Hubbard-Stratonovich transformation of the term inside brackets (Eq.~\eqref{eq:HS_transformation} with $J=1$), we have
\begin{widetext}
\begin{eqnarray}\label{eq:Lise-CanonicalZ-FieldFormulation-Derivation}
Z_N
& = & \lim_{n \rightarrow 0} \frac1n \frac{\int \prod_{\vec x} d{\vec \varphi}(\vec x) \, e^{-A} \, \left\langle \prod_{\vec x} e^{{\vec S}(\vec x) \cdot {\vec \varphi}(\vec x)} \right\rangle_{0}}{\int \prod_{\vec x} d{\vec \varphi}(\vec x) \, e^{-A}} \nonumber\\
& = & \lim_{n \rightarrow 0} \frac1n \frac{\int \prod_{\vec x} d{\vec \varphi}(\vec x) \, e^{-A} \, \prod_{\vec x} \left\langle 1 + ({\vec S}(\vec x) \cdot {\vec \varphi}(\vec x)) + \frac12 ({\vec S}(\vec x) \cdot {\vec \varphi}(\vec x))^2 \right\rangle_{0}}{\int \prod_{\vec x} d{\vec \varphi}(\vec x) \, e^{-A}} \nonumber\\
& = & \lim_{n \rightarrow 0} \frac1n \frac{\int \prod_{\vec x} d{\vec \varphi}(\vec x) \, e^{-A} \, \prod_{\vec x} \frac12 {\vec \varphi}(\vec x)^2}{\int \prod_{\vec x}
d{\vec \varphi}(\vec x) \, e^{-A}} \, ,
\end{eqnarray}
\end{widetext}
and the last line of Eq.~\eqref{eq:Lise-CanonicalZ-FieldFormulation-Derivation} is identical to Eq.~\eqref{eq:final}.
Notice that the factorization of the product in the second line of Eq.~\eqref{eq:Lise-CanonicalZ-FieldFormulation-Derivation} and the expansion up to second-order follow straightforwardly from definitions~\eqref{eq:IntroducingTraceProperties-1}-\eqref{eq:IntroducingTraceProperties-4}.
This concludes the proof.

In Sec.~\ref{sec:LatticeModel},
we will present a suitable generalization of definitions~\eqref{eq:IntroducingTraceProperties-1}-\eqref{eq:IntroducingTraceProperties-4} to treat solutions of semi-flexible polymers with bending stiffness and monomer-monomer attractive interactions for poor-solvent conditions.

\section{The many-chain field theory}\label{sec:LatticeModel}

\subsection{The model}\label{sec:LatticeModel-Definition}
We generalize here the formalism introduced in Sec.~\ref{sec:StateOfTheArt}
and we consider a system ({\it i.e.}, a solution) of semiflexible {\it linear} polymer chains with attractive interactions between non-bonded monomer pairs modelling poor-solvent conditions~\cite{Rubinstein2003}.

Chains are arranged on the same hypercubic lattice in $d$ dimensions introduced in Sec.~\ref{sec:StateOfTheArt}.
Again, lattice spacing, linear side length, volume and total number of sites are denoted, respectively, by: $a$, $L$, $V=L^d$ and $N=V/a^d$.
Chains are self- and mutually-avoiding, {\it i.e.} any two monomers -- be they from the same or different chains -- can not occupy the same lattice site.
Chain stiffness is modeled by introducing a bending energy penalty $\epsilon_a>0$ for two consecutive bonds along the same chain forming a turn (or, {\it an angle}), while attractive interactions between non-bonded monomers are accounted for by an energy reward $-\epsilon_i<0$ ($\epsilon_a, \epsilon_i>0$) for any two monomers which are separated by a unit lattice distance and either are non-consecutive if they belong to the same chain or they are on distinct chains.

\begin{figure}
\includegraphics[width=0.45\textwidth]{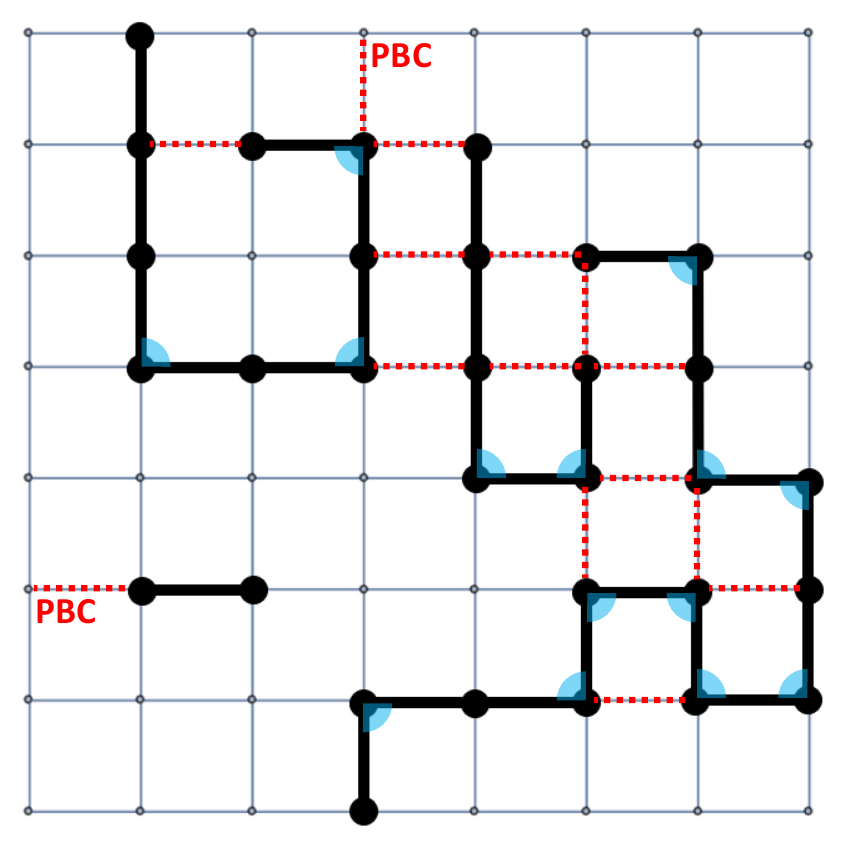}
\caption{
Schematic illustration of a particular configuration $\mathcal{C}$ on the square lattice ($d=2$).
By assuming periodic boundary conditions (PBC, see text for details), we have:
$N_c(\mathcal{C}) = 4$ chains,
$N_b(\mathcal{C}) = 27$ bonds,
$N_a(\mathcal{C}) = 14$ turns or angles (marked as blue corners),
and
$N_i(\mathcal{C}) = 15$ pairs of interacting monomers (connected by dashed red lines, with two pairs interacting through PBC.
}
\label{fig:ExampleConfiguration}
\end{figure}

For computational convenience, we work in the {\it grand canonical} ensemble where neither the number of chains nor the number of bonds are fixed and we introduce the grand canonical partition function,
\begin{equation}\label{eq:GranCanonicalZ}
Z = Z(\kappa,\eta,\epsilon_{a},\epsilon_{i}) = \sum_{ \{ {\mathcal C} \} } \, w(\kappa, \eta, \epsilon_a, \epsilon_i; {\mathcal C}) \, ,
\end{equation}
where the sum is intended over the set of all possible configurations $\{\mathcal{C}\}$ and where the thermal (Boltzmann) weight for each conformation $\mathcal C$,
\begin{equation}\label{eq:GranCanonicalZ-pi}
w(\kappa, \eta, \epsilon_a, \epsilon_i; {\mathcal C}) = \kappa^{N_b({\mathcal C})} \, \eta^{2N_c({\mathcal C})} \, e^{-\beta \epsilon_a N_a({\mathcal C}) + \beta \epsilon_i N_i({\mathcal C})} \, , 
\end{equation}
depends on:
(a)
$N_b(\mathcal{C})$, the total number of bonded monomer pairs with corresponding bond fugacity $\kappa$;
(b)
$N_c(\mathcal{C})$, the total number of chains with corresponding chain fugacity~\cite{WhyEtaSquareNote} $\eta^{2}$;
(c)
$N_a(\mathcal{C})$ and $N_i(\mathcal{C})$, respectively the total number of corners and the total number of non-bonded monomer pairs separated by one lattice distance. 
Again, $\beta=1/(k_BT)$ is the Boltzmann factor at temperature $T$ and $k_B$ is the Boltzmann constant, and we assume periodic boundary conditions through the hypercubic lattice.
Less obviously, we anticipate here and justify briefly in Sec.~\ref{sec:LatticeModel-FieldTheory}
that closed chains are ruled out in our field theory.
An example illustrating a particular configuration $\mathcal{C}$ on the square lattice ($d=2$) is shown in Fig.~\ref{fig:ExampleConfiguration}.
Notice that, per our definition, the smallest length of a single chain corresponds to $1$ lattice bond.

\subsection{Partition function and field-theoretic representation}\label{sec:LatticeModel-FieldTheory}
The central quantity of our work, the grand canonical partition function $Z$~\eqref{eq:GranCanonicalZ}, admits a field-theoretic representation.

To show it, the first point consists in devising a method ``to count'' the total number of bonds ($N_b$), chains ($N_c$), angles ($N_a$), and non-bonded monomer-monomer pairs ($N_i$) characterizing each given chain configuration $\mathcal C$.
To this purpose, we start by defining the scalar function for the configuration $\mathcal C$,
\begin{equation}
\omega_{\mathcal C}(\vec x) =
\left\{
\begin{array}{cl}
1 \, , & \text{if lattice position $\vec x$} \\
& \text{is occupied by a monomer} \\
\\
0 \, , & \text{otherwise}
\end{array}
\right.
\end{equation}
By using the matrix $\Delta(\vec x, \vec x^{\prime})$ (Eq.~\eqref{eq:def_delta}), we have
\begin{equation}\label{occnum}
N_b(\mathcal C) + N_i(\mathcal C) = \frac12 \sum_{\vec x, \vec x^{\prime}} \Delta(\vec x, \vec x^{\prime}) \, \omega_{\mathcal C}(\vec x) \, \omega_{\mathcal C}(\vec x^{\prime}) \, ,
\end{equation}
and the Hubbard-Stratonovich transformation~\cite{HubbardPRL1959,Chaikin00} of the exponential of the r.h.s of Eq.~\eqref{occnum} is equivalent to the expression containing the scalar field $\psi = \psi(\vec x)$
\begin{equation}\label{occhs}
\int \mathcal{D}\psi \, \exp \! \left[ -\frac12 \sum_{\vec x, \vec x^{\prime}} \Delta^{-1}(\vec x, \vec x^{\prime}) \, \psi(\vec{x}) \, \psi(\vec x^{\prime}) + \sum_{\vec x} \omega_{\mathcal C}(\vec x) \, \psi(\vec{x}) \right]
\end{equation}
with
\begin{equation}\label{eq:DPsi-Def}
{\mathcal D}\psi \equiv (2\pi)^{-N/2} \, (\det{\Delta})^{-1/2} \, \prod_{\vec{x}}d\psi(\vec{x}) \, .
\end{equation}
Then, at each lattice position $\vec x$ we introduce $d$ $n$-component vectors, ${\vec S}_{\mu}(\vec x) \equiv (S_{\mu}^1(\vec x), S_{\mu}^2(\vec x), ..., S_{\mu}^n(\vec x))$ with $\mu = 1, 2, ..., d$, obeying the generalized {\it trace} rules:
\begin{eqnarray}
\langle 1 \rangle_0 & = & 1 \, , \label{eq:TraceProperties-1} \\
\langle S^i_{\mu} \rangle_0 & = & 0 \, , \label{eq:TraceProperties-2} \\
\langle S^i_{\mu} S^{j}_{\nu} \rangle_0 & = & \delta_{ij}[\delta_{\mu \nu} + (1-\delta_{\mu\nu}) e^{-\beta\epsilon_a}] \, , \label{eq:TraceProperties-3} \\
\langle S^{i_1}_{\mu_1} S^{i_2}_{\mu_2} \, ... \, S^{i_k}_{\mu_k} \rangle_0 & = & 0 \, , \, \, \, \text{if $k \geq 3$} \, . \label{eq:TraceProperties-4}
\end{eqnarray}
Again, $S$-vectors on different sites are independent from each other under the trace operations just defined.

By using Eqs.~\eqref{occnum}-\eqref{eq:DPsi-Def} and the discussion in Appendix~\ref{App:Exp-BetaH-expansion},
\begin{widetext}
\begin{eqnarray}\label{eq:partitionfunction}
Z
& = & \int\mathcal{D}\psi \, e^{-\frac12 \sum_{\vec x, \vec x^{\prime}} \Delta^{-1}(\vec x, \vec x^{\prime}) \, \psi(\vec x) \, \psi(\vec x^{\prime})} \nonumber\\
& & \times \lim_{n\to 0} \bigg\langle \prod_{\vec x} \bigg\{
\bigg( 1+H(\vec x) \sum_{\mu=1}^d S_{\mu}^1 (\vec x) \bigg) \prod_{\mu = 1}^d \bigg[ 1 + h(\vec x)h(\vec x + {\hat e_{\mu}}) \, {\vec S}_{\mu}(\vec x) \cdot {\vec S}_{\mu}(\vec x + {\hat e_{\mu}}) \bigg] \bigg\} \bigg\rangle_0 \, , \nonumber\\
\end{eqnarray}
\end{widetext}
where ${\hat e_{\mu}}$ is the unit lattice vector pointing towards the ``$+\mu$''-direction introduced in Sec.~\ref{sec:StateOfTheArt}
and
\begin{eqnarray}
H(\vec{x}) & = & \frac{\eta}{1+(d-1)e^{-\beta\epsilon_a}} e^{\frac{\sqrt{\beta\epsilon_{i}}}{2}\psi(\vec x)} \, , \nonumber\\
h(\vec{x}) & = & \sqrt{\kappa}e^{-\beta\epsilon_i/2} e^{\frac{\sqrt{\beta\epsilon_i}}2 \psi(\vec x)} \, . \nonumber
\end{eqnarray}
Importantly, it must be noticed that Eq.~\eqref{eq:partitionfunction} takes into account the fact that there must be no branching points (owing to the trace definitions~\eqref{eq:TraceProperties-1}-\eqref{eq:TraceProperties-4}, any branching point gives a contribution equal to $0$) and no closed loops (from Appendix~\ref{App:Exp-BetaH-expansion},
the statistical weight of any configuration $\mathcal C$ with $k$ closed-loops is proportional to $n^k$ and so its contribution disappears in the $n \to 0$ limit).
In fact, the term of Eq.~\eqref{eq:partitionfunction} appearing under the limit of $n \to 0$ can be also written as
\begin{widetext}
\begin{eqnarray}\label{eq:partitionfunction-1}
& & \lim_{n\to 0} \left\langle \prod_{\vec x} \bigg\{ \bigg( 1+H(\vec x) \sum_{\mu=1}^d S_{\mu}^1(\vec x) \bigg) \exp\bigg[ \sum_{\mu=1}^d h(\vec x) h(\vec x + {\hat e_{\mu}}) \, {\vec S}_{\mu}(\vec{x}) \cdot \vec{S}_{\mu}(\vec x + {\hat e_{\mu}}) \bigg] \bigg\} \right\rangle_0 \nonumber\\
& = & \lim_{n\to 0} \left\langle \prod_{\vec x} \bigg( 1+H(\vec x) \sum_{\mu=1}^d S_{\mu}^1(\vec x) \bigg) \exp\bigg[ \frac12 \sum_{\vec x, \vec x^{\prime}} \sum_{\mu=1}^d \Delta_{\mu}(\vec x, \vec x^{\prime}) \, h(\vec{x}) h(\vec x^{\prime}) \, {\vec S}_{\mu}(\vec x) \cdot {\vec S}_{\mu}(\vec x^{\prime}) \bigg] \right\rangle_0 \, , \nonumber\\
\end{eqnarray}
\end{widetext}
since, again because of the trace definitions~\eqref{eq:TraceProperties-1}-\eqref{eq:TraceProperties-4}, in the expansion of the exponential all terms higher than the first are of order $n$ and they disappear in the $n \to 0$ limit and where, by analogy with the matrix $\Delta$ (Eq.~\eqref{eq:def_delta}), we have defined
\begin{equation}\label{eq:def_delta_mu}
\Delta_{\mu}(\vec x, \vec x^{\prime}) =
\left\{
\begin{array}{cl}
1 \, , & \text{if} \, \, \, |(\vec x - \vec x^{\prime}) \cdot {\hat e_{\mu}}| = a \\
0 \, , & \text{otherwise} 
\end{array}
\right. \, .
\end{equation}

For expression~\eqref{eq:partitionfunction} to be helpful, we have to remove the dependence on the $\vec S$-vectors in favor of real fields.
To this purpose, we perform a Hubbard-Stratonovich transformation~\cite{HubbardPRL1959,Chaikin00} of the exponential term of the last line of Eq.~\eqref{eq:partitionfunction-1} containing the $\vec{S}$-vectors,
\begin{widetext}
\begin{equation}\label{eq:HubbardStratonovich-PhiFields}
e^{ \frac12 \sum_{\vec x, \vec x^{\prime}} \sum_{\mu=1}^d \Delta_{\mu}(\vec x, \vec x^{\prime}) \, h(\vec{x}) h(\vec x^{\prime}) \, {\vec S}_{\mu}(\vec x) \cdot {\vec S}_{\mu}(\vec x^{\prime}) }
= \int {\mathcal D}\varphi \, e^{ -\frac12 \sum_{\vec x, \vec x^{\prime}} \sum_{\mu=1}^d \Delta_{\mu}^{-1}(\vec x, \vec x^{\prime}) \, {\vec \varphi}_{\mu}(\vec x) \cdot {\vec \varphi}_{\mu}(\vec x^{\prime}) + \sum_{\vec x} h(\vec x) \sum_{\mu=1}^d {\vec S}_{\mu}(\vec x) \cdot {\vec \varphi}_{\mu}(\vec x) } \, ,
\end{equation}
\end{widetext}
where we have introduced
the $d$ real {\it vector} fields
${\vec \varphi}_{\mu}(\vec x)$ ($\mu = 1, \dots, d)$
with
${\vec \varphi}_{\mu}(\vec x) \equiv (\varphi_\mu^1(\vec x), \varphi_\mu^2(\vec x), ..., \varphi_\mu^n(\vec x))$
and the corresponding measure (see Eq.~\eqref{eq:DPsi-Def}, for analogy)
\begin{equation}\label{eq:DPhi-Def}
{\mathcal D}\varphi \equiv (2\pi)^{-ndN/2} \prod_{\mu=1}^d \left( \det{\Delta_{\mu}} \right)^{-n/2} \prod_{\vec x} \prod_{\mu=1}^d d{\vec \varphi}_{\mu}(\vec x) \, .
\end{equation}

Finally by
(i)
inserting Eq.~\eqref{eq:HubbardStratonovich-PhiFields} into Eq.~\eqref{eq:partitionfunction} via Eq.~\eqref{eq:partitionfunction-1};
(ii)
Taylor-expanding the term containing the $\vec \varphi_{\mu}$-fields;
(iii)
applying the trace definitions~\eqref{eq:TraceProperties-1}-\eqref{eq:TraceProperties-4}
and
(iv)
noticing that the first two terms in Eq.~\eqref{eq:DPhi-Def} give $=1$ in the limit $n\to 0$, one can show that, up to an unimportant multiplicative constant,
\begin{widetext}
\begin{equation}\label{eq:fullz}
Z = \lim_{n \to 0} \int \prod_{\vec x} d\psi(\vec x) \int \prod_{\vec x} \prod_{\mu=1}^d d\vec{\varphi}_{\mu}(\vec x) \exp\left\{ -A[\{\psi\}] - \sum_{\mu=1}^d A_{\mu}[\{\vec{\varphi}_{\mu}\}] + \sum_{\vec x} \ln \left[ 1 + e^{\sqrt{\beta \epsilon_i} \psi(\vec x)} B[\{\vec{\varphi}_{\mu}(\vec x)\}] \right] \right\} \, ,
\end{equation}
\end{widetext}
where we have defined the following functionals:
\begin{widetext}
\begin{eqnarray}
A[\{\psi\}]
& = & \frac12 \sum_{\vec x, \vec x^{\prime}} \Delta^{-1} \! (\vec x, \vec x^{\prime}) \, \psi(\vec x) \, \psi(\vec x^{\prime}) \, , \label{eq:Apsi} \\
A_{\mu}[\{\vec{\varphi}_{\mu}\}]
& = & \frac12 \sum_{\vec x, \vec x^{\prime}} \Delta_{\mu}^{-1} \! (\vec x, \vec x^{\prime}) \, {\vec \varphi}_{\mu}(\vec{x}) \cdot \vec{\varphi}_{\mu}(\vec x^{\prime}) \, , \label{eq:Amu} \\
B[\{\vec{\varphi}_{\mu}(\vec x)\}]
& = & \frac{\kappa e^{-\beta\epsilon_i}}2 \left[ (1-e^{-\beta\epsilon_a}) \sum_{\mu = 1}^d |{\vec \varphi}_{\mu}({\vec x})|^{2} + e^{-\beta\epsilon_a} \left( \sum_{\mu=1}^d {\vec \varphi}_{\mu}(\vec x) \right)^2 \right] + \sqrt{\kappa}\eta e^{-\beta\epsilon_{i}/2}\sum_{\mu = 1}^{d}\varphi_{\mu}^{1}(\vec x) \, . \label{eq:Bphimu}
\end{eqnarray}
\end{widetext}
Importantly, notice the explicit presence of $\varphi_{\mu}^1(\vec{x})$ in Eq.~\eqref{eq:Bphimu}.
This is a direct consequence of the fact that, in order to describe a system of multiple chains through the $O(n \to 0)$ formalism, it suffices to introduce an external magnetic field in the spin Hamiltonian~\cite{desCloizeaux1975,deGennes1979}.
This field can pick any arbitrary direction: in our derivation, we have chosen the direction with $\mu=1$.
As a validation of Eqs.~\eqref{eq:fullz}-\eqref{eq:Bphimu}, we report that, in the ``single-chain'' limit $\eta\rightarrow0$, we get back the original result by Doniach {\it et al.}~\cite{Doniach1996} for a single semiflexible chain with non-bonded attractive interactions in the presence of lattice vacancies.

\section{Mean-field solution: saddle-point approximation}\label{sec:LatticeModel-MFApprox}
The exact grand canonical partition function $Z$ (Eq.~\eqref{eq:fullz}) is the central result of this work.
As for the single chain case a direct evaluation of $Z$ is not feasible but the field theoretical formulation (Eq.~\eqref{eq:fullz}) is very suitable for its mean field (MF) treatment~\cite{Orland1985,Bascle1992,Doniach1996}.

We start by differentiating the exponential in Eq.~\eqref{eq:fullz} with respect to $\varphi^{i}_{\mu}(\vec x)$ and $\psi(\vec x)$ and set the obtained expressions equal to $0$ in order to get the stationary solution.
We further take the solutions to be homogeneous assuming translational invariance and break the $O(n)$ symmetry of the vector field so that
\begin{eqnarray}
\vec{\varphi}_{\mu}(\vec{x})
& = & (\varphi, 0, \dots, 0) \label{eq:SearchPhiSol} \, , \\
\psi(\vec{x})
& = & \psi \label{eq:SearchPsiSol} \, ,
\end{eqnarray}
for every $\vec{x}$ and every $\mu$, thus obtaining 
\begin{eqnarray}
\frac{\varphi}2
& = & \frac{e^{\sqrt{\beta\epsilon_i}\psi}(\frac{\kappa e^{-\beta\epsilon_i} q(\beta)}2 \, \varphi + \sqrt{\kappa} \, \eta \, e^{-\beta\epsilon_i/2})}{1+e^{\sqrt{\beta\epsilon_i}\psi}d\bigg(\frac{\kappa e^{-\beta\epsilon_i} q(\beta)}4 \, \varphi^2 + \sqrt{\kappa} \, \eta \, e^{-\beta\epsilon_i/2} \, \varphi \bigg)} \, , \nonumber\\
& & \label{eq:phisp} \\
\frac{\psi}{2d}
& = & \frac{\sqrt{\beta\epsilon_i} \, e^{\sqrt{\beta\epsilon_i}\psi}d\bigg(\frac{\kappa e^{-\beta\epsilon_i} q(\beta)}4 \, \varphi^2 + \sqrt{\kappa} \, \eta \, e^{-\beta\epsilon_i/2} \, \varphi\bigg)}{1+e^{\sqrt{\beta\epsilon_i}\psi}d\bigg(\frac{\kappa e^{-\beta\epsilon_i} q(\beta)}4 \, \varphi^{2} + \sqrt{\kappa} \, \eta \, e^{-\beta\epsilon_i/2} \, \varphi\bigg)} \, , \nonumber\\
& & \label{eq:psisp}
\end{eqnarray}
where~\cite{Doniach1996}
$q(\beta) = 2+2(d-1)e^{-\beta\epsilon_a}$.
In terms of the solutions~\cite{OnPhiPsiNotation} 
$\varphi = \varphi(\kappa, \eta, \epsilon_i, \epsilon_a)$
and
$\psi = \psi(\kappa, \eta, \epsilon_i, \epsilon_a)$
of the MF Eqs.~\eqref{eq:phisp} and~\eqref{eq:psisp}, the grand potential per lattice site (up to unimportant additive constants) reads
\begin{widetext}
\begin{equation}\label{eq:completefreeene}
\beta \Omega(\kappa,\eta, \epsilon_{a}, \epsilon_{i}) = \frac{\psi^2}{4d} +\frac{d\varphi^2}4 - \ln \bigg[ 1 + d \, e^{\sqrt{\beta\epsilon_i} \, \psi} \, \bigg( \frac{\kappa e^{-\beta\epsilon_i} q(\beta)}4 \, \varphi^2 + \sqrt{\kappa} \, \eta \, e^{-\beta\epsilon_i/2} \, \varphi \bigg) \bigg] \, .
\end{equation}
\end{widetext}
Notice that with the ansatz~\eqref{eq:SearchPhiSol} and~\eqref{eq:SearchPsiSol} every dependence upon $n$ disappears, and thus the limit $n \to 0$ is trivial.

On setting $\eta = 0$, Eqs.~\eqref{eq:phisp} and~\eqref{eq:psisp} reduce to the ones obtained in~\cite{Doniach1996} for the single chain model.
In the following, we will thus solve the saddle-point equations in the case $\eta > 0$ that will be then compared with Monte Carlo simulations (Sec.~\ref{sec:LatticeModel-MC})
in Sec.~\ref{sec:MFvsMC-Results}.

\section{Monte Carlo simulations}\label{sec:LatticeModel-MC}
In order to check the validity of the MF approximation as well as to assess its limits, we have performed Metropolis~\cite{Metropolis1953} Grand Canonical Monte Carlo (GCMC) computer simulations of the lattice model (Sec.~\ref{sec:LatticeModel})
on the three-dimensional cubic lattice. 
Essentially, the goal of the GCMC simulations is to obtain a representative sample of polymer configurations in agreement with the grand canonical partition function~\eqref{eq:GranCanonicalZ}.

The implementation of our algorithm is relatively straightforward, and it works as the following.
As explained in Sec.~\ref{sec:LatticeModel},
the Boltzmann weight $w$ (see Eq.~\eqref{eq:GranCanonicalZ-pi}) of each polymer configuration $\mathcal C$ in the ensemble is a function of the total number of bonds ($N_b({\mathcal C})$), distinct chains ($N_c({\mathcal C})$), turns ($N_a({\mathcal C})$) and pairs of non-bonded nearest-neighbor monomers ($N_i({\mathcal C})$).
Therefore, at each MC step one single bond is randomly inserted in or removed from the lattice, provoking a change of the configuration ${\mathcal C}_0$ to the configuration ${\mathcal C}_1$.
In order to enforce the condition of {\it detailed balance}, we accept~\cite{Metropolis1953} the new conformation with probability given by the expression:
\begin{equation}\label{eq:accRatioIdeal}
\mbox{acc}({\mathcal C}_0 \rightarrow {\mathcal C}_1) =
\left\{
\begin{array}{cc}
\min \left\{ 1, \frac{d}{\phi_{b,1}} \, \frac{w(\kappa, \eta, \epsilon_a, \epsilon_i; \, {\mathcal C_1})}{w(\kappa, \eta, \epsilon_a, \epsilon_i; \, {\mathcal C}_0)} \right\} & \mbox{(bond inserted)} \\
\\
\min \left\{ 1, \frac{\phi_{b,0}}d \, \frac{w(\kappa, \eta, \epsilon_a, \epsilon_i; \, {\mathcal C_1})}{w(\kappa, \eta, \epsilon_a, \epsilon_i; \, {\mathcal C}_0)} \right\} & \mbox{(bond removed)}
\end{array}
\right. \, ,
\end{equation}
where $\phi_{b, 0}$ (respectively, $\phi_{b, 1}$) is the bond density (see Eq.~\eqref{eq:phibSol}) of the configuration ${\mathcal C}_0$ (resp., ${\mathcal C}_1$).
Whenever the insertion of a new bond leads to a forbidden configuration ({\it e.g.}, for the presence of branching points or closed loops), the move is automatically discarded.

In order to check for finite-size effects, we have performed preliminary calculations and compared corresponding results for lattice sizes $L/a=4,8,16$.
This analysis indicates that for $L/a\geq 8$ the results do not change significantly with the lattice size and hence we will fix $L/a=8$ in all calculations henceforth.
This  guarantees a good compromise between computational efficiency and accuracy.

For each chosen pair of $(\kappa, \eta)$ values, we have led a simulation run consisting of $10^7$ MC steps.
Then, for each GCMC trajectory, a standard block analysis procedure~\cite{Newman1999} has been carried out in order to estimate uncertainties on the considered physical observables.
Every trajectory has been checked individually in order to make sure that all the curves obtained from corresponding block analyses and representing the MC-time evolution of the distinct quantities have completely equilibrated.
This procedure has been applied to the number of bonds, the number of chains and the internal energy of the system.

\section{Mean-field solution {\it vs.} Monte Carlo simulations}\label{sec:MFvsMC-Results}
From our mean-field estimate of the grand potential, Eq.~\eqref{eq:completefreeene}, we can compute:
\begin{itemize}
\item
The {\it bond} density,
\begin{equation}\label{eq:phibSol}
\phi_b \equiv \frac{\langle N_b \rangle}N = -\beta\kappa\frac{\partial \Omega(\kappa, \eta, \epsilon_i, \epsilon_a)}{\partial \kappa} = \frac{d}4 \, \varphi^2 \, ;
\end{equation}
\item
The {\it chain} density,
\begin{eqnarray}\label{eq:phicSol}
\phi_c
& \equiv & \frac{\langle N_c \rangle}N = -\frac{\beta\eta}2 \, \frac{\partial \Omega(\kappa,\eta, \epsilon_{i},\epsilon_{a})}{\partial \eta} \nonumber\\
& = & \frac{d}2 \, \frac{e^{\sqrt{\beta\epsilon_i} \, \psi}\sqrt{\kappa} \, \eta \, e^{-\beta\epsilon_i/2} \, \varphi}{1 + d \, e^{\sqrt{\beta\epsilon_i}\psi} \bigg(\frac{\kappa e^{-\beta\epsilon_i} q(\beta)}4 \, \varphi^2 + \sqrt{\kappa} \, \eta \, e^{-\beta\epsilon_i/2} \, \varphi \bigg)} \, ; \nonumber\\
\end{eqnarray}
\item
The {\it total} monomer density,
\begin{equation}\label{eq:MonomerPhi}
\phi \equiv \phi_b + \phi_c = \frac1{\sqrt{\beta\epsilon_i}} \, \frac{\psi}{2d} \, .
\end{equation}
\end{itemize}
In the rest of this Section, we specialize the saddle-point equations~\eqref{eq:phisp} and~\eqref{eq:psisp} to various particular cases and compare the corresponding results to Monte Carlo simulations 
in $d=3$. 

\subsection{Case $\epsilon_a=\epsilon_i=0$}\label{sec:Entropy1Stiffness0Energy0}
The simplest case to be considered is the case of non-interacting (still non-overlapping) flexible chains with no bending penalty nor monomer-monomer attractive interactions.
In spite its simplicity, this case proves to be rather instructive. In this case Eqs.~\eqref{eq:phisp} and~\eqref{eq:psisp} read
\begin{eqnarray}
\frac{\varphi}2
& = & \frac{\kappa d \, \varphi + \sqrt{\kappa} \, \eta}{1+d\bigg(\frac{\kappa d}2 \, \varphi^{2} + \sqrt{\kappa} \, \eta \, \varphi\bigg)} \, , \label{eq:phisp-Entropy1Stiffness0Energy0} \\
\frac{\psi}{2d}
& = & 0 \, . \label{eq:psisp-Entropy1Stiffness0Energy0}
\end{eqnarray}
The only relevant field is thus $\varphi$ and, since it satisfies the simple cubic equation~\eqref{eq:phisp-Entropy1Stiffness0Energy0}, in principle three (real) solutions can be possible. However two additional constraints identify the only acceptable solution.
First, the argument of the logarithm in Eq.~\eqref{eq:completefreeene} ({\it i.e.}, the denominator in Eq.~\eqref{eq:phisp-Entropy1Stiffness0Energy0}) must be strictly $>0$. Second,
the chain density (Eq.~\eqref{eq:phicSol}) must satisfy the inequality $0 \leq \phi_c \leq 1/2$.
In Appendix~\ref{App:SolutionCubic}
we provide evidence that, for every $\kappa \geq 0$ and $\eta > 0$, there exists one and only one such solution $\varphi >0$ which is a {\it continuous} function of the parameters.
This solution can then be inserted in Eqs.~\eqref{eq:phibSol} and~\eqref{eq:phicSol} to obtain the bond and chain density, $\phi_b$ and $\phi_c$.

\begin{figure}
\includegraphics[width=0.45\textwidth]{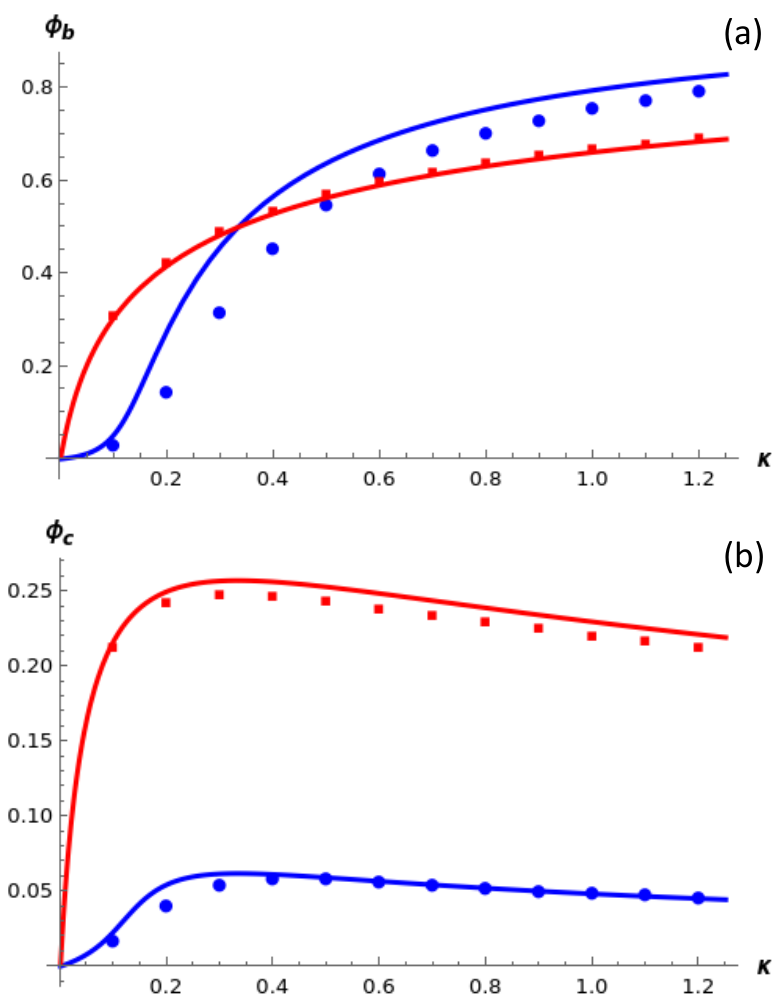}
\caption{
$\epsilon_a=\epsilon_i=0$.
Bond density $\phi_b$ (a) and chain density $\phi_c$ (b) as a function of the bond fugacity $\kappa$ and for chain fugacities $\eta = 0.2$ (blue) and $\eta = 1.5$ (red).
Solid lines and symbols are, respectively, for the MF solution and the GCMC computer simulations.
}
\label{fig:bond_e+chain_e}
\end{figure}

MF calculations for the bond and chain density, $\phi_b$ and $\phi_c$, as a function of the bond fugacity $\kappa$ and for two representative chain fugacities $\eta=0.2$ (small) and $\eta=1.5$ (large) are shown as solid lines in Fig.~\ref{fig:bond_e+chain_e} (panels (a) and (b), respectively) and compared to corresponding GCMC simulations (symbols).
The nearly quantitative agreement between the MF calculations and the GCMC simulations is remarkable, thus validating our MF approach.

One striking feature of the bond density curves (see Fig.~\ref{fig:bond_e+chain_e}(a)) is that they intersect at a certain $\kappa = \kappa^\ast$, such that $\phi_b(\kappa^\ast) \simeq 0.5$.
Although odd at first sight, this behaviour can be simply rationalized as the following.
When $\phi_b < 0.5$ it is likely that the insertion of a new bond will also lead to the creation of a new chain.
Thus, for $\kappa < \kappa^\ast$, the bond density increases faster for larger values of $\eta$ (red curve in Fig.~\ref{fig:bond_e+chain_e}(a)) than for smaller values of $\eta$ (blue curve) because configurations with a larger number of chains are more favoured.
Conversely, when $\phi_b > 0.5$ once a new bond is inserted it will link two different chains, thus reducing their total number.
Under these conditions, for $\kappa > \kappa^\ast$ the bond density increases faster for {\it smaller} values of $\eta$ (blue curve) than for larger ones (red curve).
A further support to this interpretation also stems by the fact that the chain density (Fig.~\ref{fig:bond_e+chain_e}(b)) has a maximum at $\kappa \simeq \kappa^\ast$.

It also proves instructive to derive simple analytical expressions for $\phi_b$ and $\phi_c$ in the limit $\kappa \gg 1$.
In this regime, Eq.~\eqref{eq:phisp-Entropy1Stiffness0Energy0} depends only on one parameter, namely the ratio $\eta/\sqrt{\kappa}$, and has the following solutions
\begin{equation}
\varphi \simeq
\left\{
\begin{array}{cc}
\sqrt{\frac{4}{d}}-\frac{1}{2d}\frac{\eta}{\sqrt{\kappa}} \, , & \text{if} \, \, \,  \frac{\eta}{\sqrt{\kappa}} \ll 1 \\
\\
\sqrt{\frac{2}{d}} + \sqrt{\frac{d}{8}}\frac{\sqrt{\kappa}}{\eta} \, , & \text{if} \, \, \, \frac{\eta}{\sqrt{\kappa}} \gg 1
\end{array}
\right.
\end{equation}
By plugging these results into the expressions for $\phi_b$ (Eq.~\eqref{eq:phibSol}) and $\phi_c$ (Eq.~\eqref{eq:phicSol}) we find
\begin{equation}
\phi_{b} \simeq
\left\{
\begin{array}{cc}
1-\frac{1}{2\sqrt{d}}\frac{\eta}{\sqrt{\kappa}} \, , & \text{if} \, \, \,  \frac{\eta}{\sqrt{\kappa}} \ll 1 \\
\\
\frac12 + \sqrt{\frac{d}{8}}\frac{\sqrt{\kappa}}{\eta} \, , & \text{if} \, \, \, \frac{\eta}{\sqrt{\kappa}} \gg 1
\end{array}
\right.
\end{equation}
and
\begin{equation}
\phi_{c} \simeq
\left\{
\begin{array}{cc}
\frac{1}{2\sqrt{d}}\frac{\eta}{\sqrt{\kappa}} \, , & \text{if} \, \, \, \frac{\eta}{\sqrt{\kappa}} \ll 1 \\
\\
\frac12 - \sqrt{\frac{d}{8}}\frac{\sqrt{\kappa}}{\eta} \, , & \text{if} \, \, \, \frac{\eta}{\sqrt{\kappa}} \gg 1 
\end{array}
\right.
\end{equation}
For $\kappa \gg 1$ the monomer density $\phi = \phi_b + \phi_c$ is always $\simeq 1$.
However, there are two different scenarios: if $\eta/\sqrt{\kappa} \ll 1$ then $\phi_b/\phi_c \gg 1$, {\it i.e.} the number of chains is very low, but on average they are very long. 
On the other hand, if $\eta/\sqrt{\kappa} \gg 1$ then $\phi_b/\phi_c \simeq 1$, that is the number of chains is large but they are all essentially formed by one single bond, in agreement with previous interpretation.

\subsection{Case $\epsilon_a>0$, $\epsilon_i=0$}\label{sec:Entropy1Stiffness1Energy0}
With only the contribution of the bending penalty but no monomer-monomer attractive interactions, Eqs.~\eqref{eq:phisp} and~\eqref{eq:psisp} read
\begin{eqnarray}
\frac{\varphi}2
& = & \frac{\frac{\kappa q(\beta)}2 \, \varphi + \sqrt{\kappa}\eta}{1+d\bigg(\frac{\kappa q(\beta)}4 \varphi^{2} + \sqrt{\kappa}\eta \, \varphi\bigg)} \, , \label{eq:phisp-Entropy1Stiffness1Energy0} \\
\frac{\psi}{2d}
& = & 0 \, . \label{eq:psisp-Entropy1Stiffness1Energy0}
\end{eqnarray}
It is easy to see that this is the same situation of Sec.~\ref{sec:Entropy1Stiffness0Energy0}
with renormalized fugacities $\kappa \rightarrow \kappa q(\beta) / 2d$ and $\eta \rightarrow \eta\sqrt{\frac{2d}{q(\beta)}}$.
Since $2 \leq q(\beta) \leq 2d$, introducing a non-zero bending stiffness leads ultimately to a lower effective bond fugacity and a larger effective chain fugacity.
Hence, as seen in Sec.~\ref{sec:Entropy1Stiffness0Energy0},
again we have only one acceptable solution which is a continuous function of the parameters.

\begin{figure}
\includegraphics[width=0.45\textwidth]{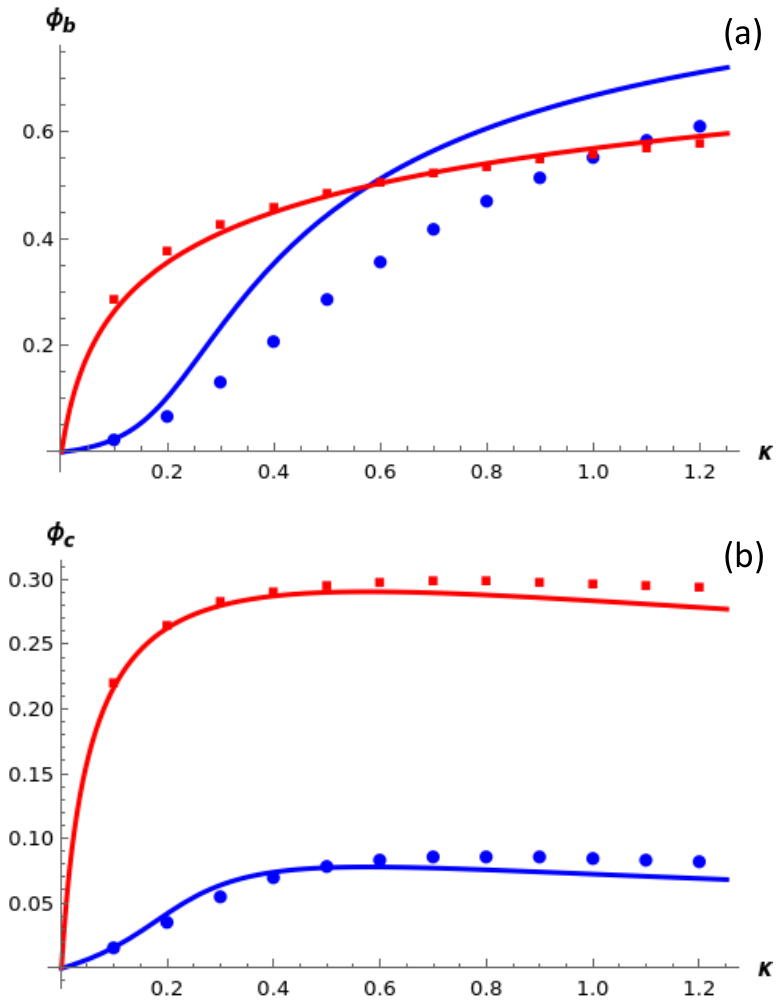}
\caption{
$\epsilon_a / k_B T = 1 \, , \epsilon_i=0$.
Bond density $\phi_b$ (a) and chain density $\phi_c$ (b) as a function of the bond fugacity $\kappa$ and for chain fugacities $\eta = 0.2$ (blue) and $\eta = 1.5$ (red).
Solid lines and symbols are as in Fig.~\ref{fig:bond_e+chain_e}.
}
\label{fig:bond_eb+chain_eb}
\end{figure}

By fixing the bending stiffness to the convenient value $\epsilon_a = k_B T$, MF calculations for the bond and chain density, $\phi_b$ and $\phi_c$, as a function of the bond fugacity $\kappa$ and for chain fugacities $\eta=0.2$ and $\eta=1.5$ are shown as solid lines in Fig.~\ref{fig:bond_eb+chain_eb} (panels (a) and (b), respectively) and compared to corresponding GCMC simulations (symbols).
As in previous Sec.~\ref{sec:Entropy1Stiffness0Energy0},
the agreement between MF calculation and GCMC simulations is remarkable. 

\begin{figure}
\includegraphics[width=0.45\textwidth]{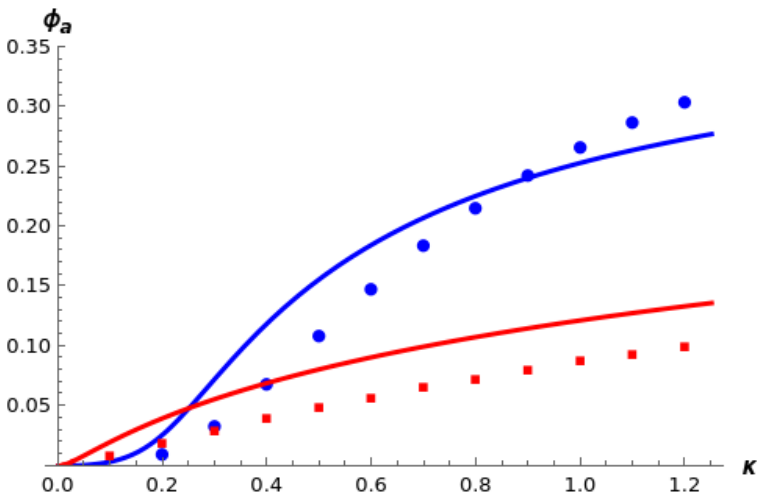}
\caption{
$\epsilon_a / k_B T = 1 \, , \epsilon_i=0$.
Angle density $\phi_a$ as a function of the bond fugacity $\kappa$ and for chain fugacities $\eta = 0.2$ (blue) and $\eta = 1.5$ (red).
Solid lines and symbols are as in Fig.~\ref{fig:bond_e+chain_e}.
}
\label{fig:angle_eb}
\end{figure}

In principle, here one would have expected an isotropic-to-nematic transition which would be observable as a singularity in the average ``angle'' density, $\phi_a$.
However, due to the fact that the only contribution of the bending stiffness is to renormalize the fugacities, our MF treatment does not display the appearance of such transition.
This is made evident in Fig.~\ref{fig:angle_eb} where we show the average angle density as obtained in the MF approximation (solid lines, see formula~\eqref{eq:AngleDensity}) in favorable comparison to the results of GCMC simulations (symbols).

\subsection{Case $\epsilon_a=0$, $\epsilon_i>0$}\label{sec:Entropy1Stiffness0Energy1}
This is the complementary case of previous one, where monomer-monomer attraction is present but there is no bending penalty.
In this case, Eqs.~\eqref{eq:phisp} and~\eqref{eq:psisp} read
\begin{eqnarray}
\frac{\varphi}2
& = & \frac{e^{\sqrt{\beta\epsilon_i}\psi}(\kappa e^{-\beta\epsilon_i}d \, \varphi + \sqrt{\kappa} \, \eta \, e^{-\beta\epsilon_i/2})}{1+e^{\sqrt{\beta\epsilon_i}\psi}d\bigg(\frac{\kappa e^{-\beta\epsilon_i} d}2 \, \varphi^{2} + \sqrt{\kappa} \, \eta \, e^{-\beta\epsilon_i/2} \, \varphi\bigg)} \, , \label{eq:phisp-Entropy1Stiffness0Energy1} \nonumber\\
& & \\
\frac{\psi}{2d}
& = & \frac{\sqrt{\beta\epsilon_i} \, e^{\sqrt{\beta\epsilon_i}\psi}d\bigg(\frac{\kappa e^{-\beta\epsilon_i} d}2 \, \varphi^2 + \sqrt{\kappa} \, \eta \, e^{-\beta\epsilon_i/2} \, \varphi\bigg)}{1+e^{\sqrt{\beta\epsilon_i}\psi}d\bigg(\frac{\kappa e^{-\beta\epsilon_i} d}2 \, \varphi^2 + \sqrt{\kappa} \, \eta \, e^{-\beta\epsilon_i/2} \, \varphi\bigg)} \, . \label{eq:psisp-Entropy1Stiffness0Energy1} \nonumber\\
\end{eqnarray}
which do not admit a simple closed solution for $\psi$ as in the previous cases.
Hence these equation have to be solved numerically with the two constraints discussed in Sec.~\ref{sec:Entropy1Stiffness0Energy0}
plus the condition $0\leq \phi \leq 1$, that implies $0 \leq \psi \leq 2d\sqrt{\beta\epsilon_i}$, for the total monomer density $\phi$ (Eq.~\eqref{eq:MonomerPhi}).
Interestingly, 
in this case one finds that there are multiple acceptable solutions for this system, and hence the most stable solution corresponds to that minimizing the grand potential $\Omega$ (Eq.~\eqref{eq:completefreeene}).
In particular, 
this leads to the appearance of discontinuities in $\phi_b$ and $\phi_c$ 
(see Appendix~\ref{App:ShowPhaseTransition} for details). 

\begin{figure}
\includegraphics[width=0.45\textwidth]{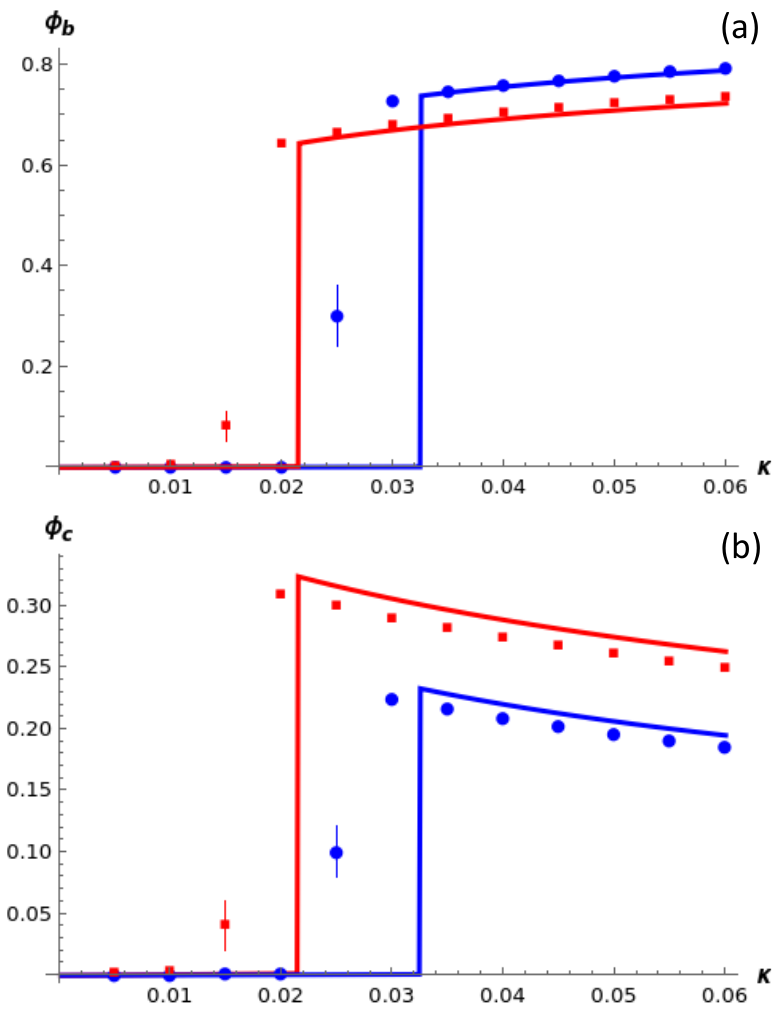}
\caption{
$\epsilon_a = 0 \, , \epsilon_i / k_B T = 1$.
Bond density $\phi_b$ (a) and chain density $\phi_c$ (b) as a function of the bond fugacity $\kappa$ and for chain fugacities $\eta = 0.15$ (blue) and $\eta = 0.25$ (red). 
The discontinuity predicted by MF calculations (lines) is confirmed by GCMC simulations (symbols), yet the critical value of $\kappa$ is slightly different between theory and simulations.
}
\label{fig:bond_ei+chain_ei}
\end{figure}

In fact, by fixing the monomer-monomer interaction to the convenient value $\epsilon_i = k_B T$, MF calculations for the bond and chain density, $\phi_b$ and $\phi_c$, as a function of the bond fugacity $\kappa$ and for chain fugacities $\eta=0.15$ and $\eta=0.25$ are shown as solid lines in Fig.~\ref{fig:bond_ei+chain_ei} (panels (a) and (b), respectively) and compared to corresponding GCMC simulations (symbols). 
In both panels one can easily distinguish two different phases, with both densities acting as order parameters.
In the first (gas-like) phase, the total monomer density $\phi = \phi_b + \phi_c$ is close to 0 and $\phi_b/\phi_c \simeq 1$.
This is valid up to some critical value $\kappa_{\rm cr}$ above which $\phi$ is close to 1 and $\phi_b/\phi_c > 1$ (liquid-like phase).

By varying systematically the parameters $\beta\epsilon_i$ and $\eta$, we have extracted each corresponding critical value $\kappa_{\rm cr}$ through the numerical solution of the coupled Eqs.~\eqref{eq:phisp-Entropy1Stiffness0Energy1} and~\eqref{eq:psisp-Entropy1Stiffness0Energy1}.
Three illustrative ``coexistence'' lines corresponding to the values $\beta\epsilon_i = 0.75$, $1.00$ and $1.25$ are shown in panel (a) of Fig.~\ref{fig:PhaseDiagram} (solid lines).
Similarly (panel (b), filled symbols), we have determined the values of the total monomer density $\phi$ (Eq.~\eqref{eq:MonomerPhi}) at the coexistence by varying $\beta\epsilon_i$ systematically and for the two representative fugacity values $\eta=0.1$ and $\eta=0.3$.
We will discuss these results in full fledged way in Sec.~\ref{sec:Concls}.

\subsection{Case $\epsilon_a>0$, $\epsilon_i>0$}\label{sec:Entropy1Stiffness1Energy1}
When both the attractive monomer-monomer interaction and bending stiffness are non-zero, we need to solve the complete system of Eqs.~\eqref{eq:phisp} and~\eqref{eq:psisp}.
Again, the strategy is the same as in Sec.~\ref{sec:Entropy1Stiffness1Energy0}:
renormalizing the fugacities ($\kappa \rightarrow \kappa q(\beta) / 2d$ and $\eta \rightarrow \eta\sqrt{\frac{2d}{q(\beta)}}$) to absorb the terms accounting for the bending stiffness and obtain a system of equations which is equivalent to that presented in Sec.~\ref{sec:Entropy1Stiffness0Energy1}.
The general behavior of the solutions will thus be exactly the same 
(and, so, the discontinuities in $\phi_b$ and $\phi_c$) 
as for a system with no bending stiffness, the only difference being in the changing of the coexistence line between the two phases in the ($\kappa$,$\eta$)-plane, depending on the value of $\epsilon_a$.

\begin{figure}
\includegraphics[width=0.45\textwidth]{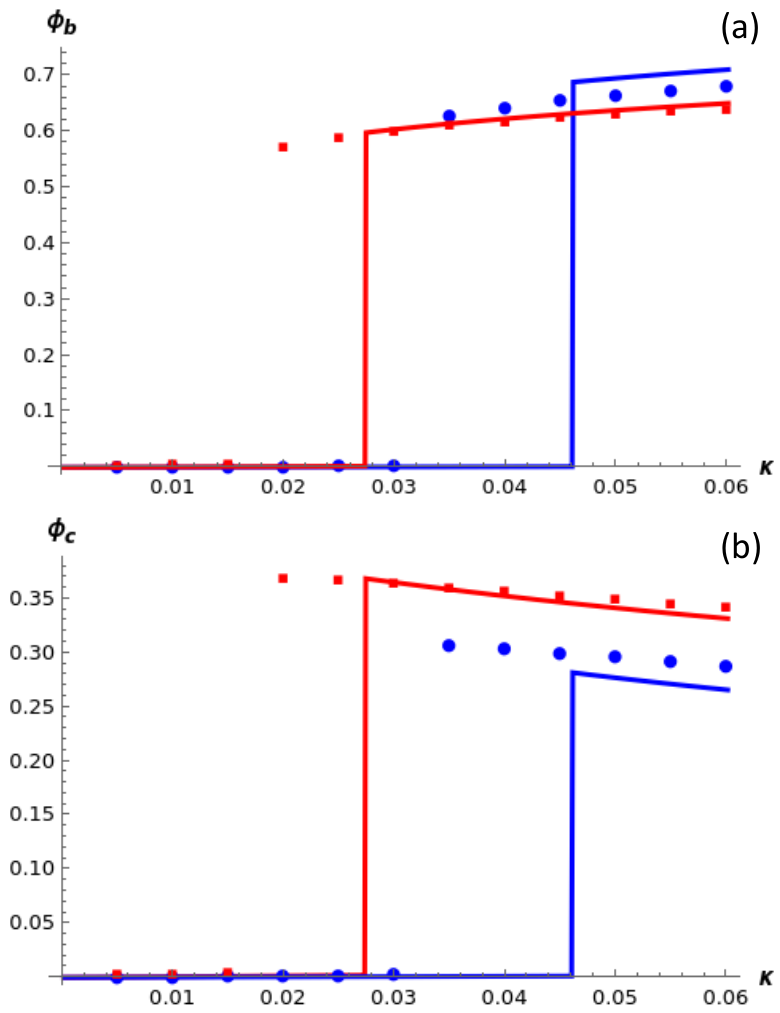}
\caption{
$\epsilon_a /k_B T = \epsilon_i / k_B T = 1$.
Bond density $\phi_b$ (a) and chain density $\phi_c$ (b) as a function of the bond fugacity $\kappa$ and for chain fugacities $\eta = 0.15$ (blue) and $\eta = 0.25$ (red).
As in the case with no bending stiffness (Fig.~\ref{fig:bond_ei+chain_ei}), the discontinuity predicted by MF calculations (lines) is apparent and its presence is confirmed by GCMC simulations (symbols).
Again, the critical value for $\kappa$ is quantitatively different between theory and simulations. 
}
\label{fig:bond_ebi+chain_ebi}
\end{figure}

\begin{figure}
\includegraphics[width=0.45\textwidth]{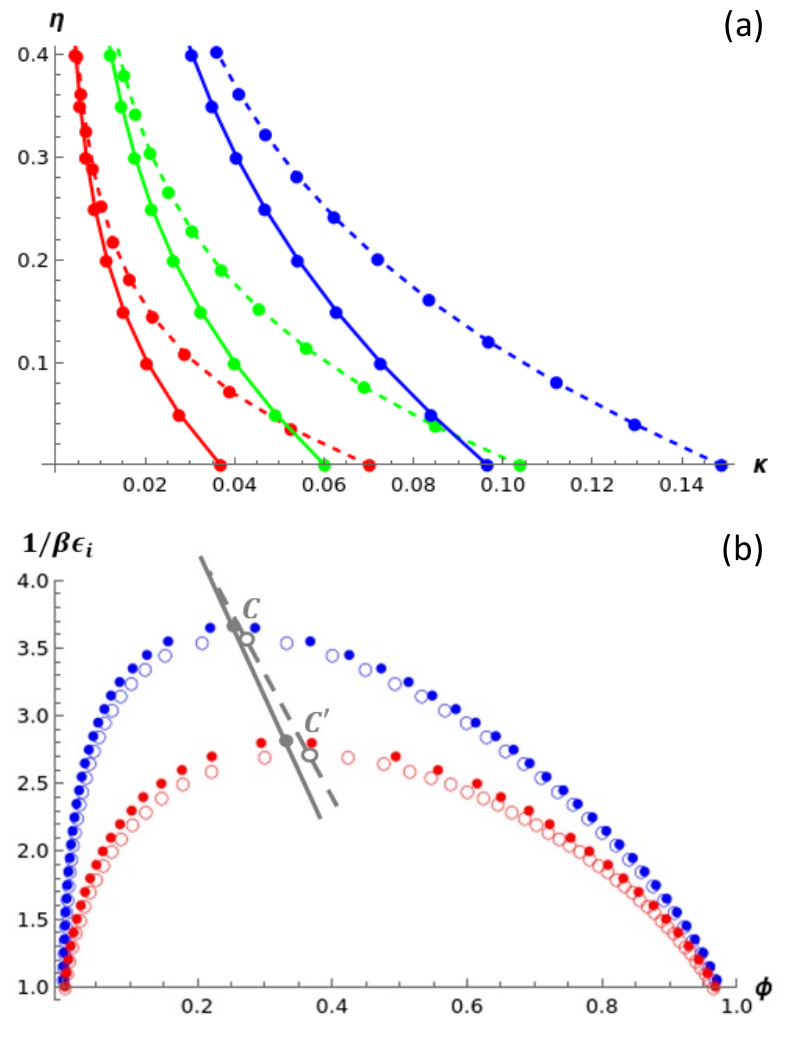}
\caption{
(a)
Coexistence lines between the gas phase and the liquid phase in the ($\kappa$, $\eta$)-plane.
Solid lines correspond to $\epsilon_a = 0$, whereas dashed lines are for $\epsilon_a/\epsilon_i = 1$.
Lines colors blue, green and red are for $\epsilon_i / k_BT = 0.75, 1.00$ and $1.25$, respectively. 
Below the coexistence line the system is in the gas phase ($\phi \simeq 0$), whereas above the coexistence line it is found in the liquid phase ($\phi \simeq 1$).
(b)
Gas-liquid transition in the ($\phi$, $1/(\beta\epsilon_i)$)-plane.
Above the critical point, the system is in a single homogeneous gas phase.
Below the critical point, the system phase-separate in a gas phase coexisting with a liquid phase, and the figure display the coexistence (binodal) line.
Symbols colors blue and red are for $\eta = 0.1$ and $\eta = 0.3$, respectively.
Filled symbols correspond to $\epsilon_a = 0$, whereas empty symbols are for $\epsilon_a/ \epsilon_i = 1$.
For each data set the relative grey symbol marks the value of the corresponding ``critical'' temperature (letters ``$C$'' and ``$C'$''), with grey lines used for guiding the eye.
}
\label{fig:PhaseDiagram}
\end{figure}

By fixing again the bending stiffness and the monomer-monomer interaction to the values $\epsilon_a = \epsilon_i = k_B T$, MF calculations for the bond and chain density, $\phi_b$ and $\phi_c$, as a function of the bond fugacity $\kappa$ and for chain fugacities $\eta=0.15$ and $\eta=0.25$ are shown as solid lines in Fig.~\ref{fig:bond_ebi+chain_ebi} (panels (a) and (b), respectively) and compared to corresponding GCMC simulations (symbols).
As expected, GCMC simulations confirm MF calculations and we can distinguish once again the gas ($\phi = \phi_b + \phi_c \approx 0$) and liquid ($\phi \approx 1$) phases.
Finally, analogously to Sec.~\ref{sec:Entropy1Stiffness0Energy1},
we produce examples of coexistence and gas-liquid transition lines, see panel (a) (dashed lines, different colors are for different $\epsilon_i$ (see caption)) and panel (b) (empty symbols, different colors are for different $\eta$ (see caption)) of Fig.~\ref{fig:PhaseDiagram}.

\section{Connections to the Flory-Huggins theory of mixing}\label{sec:FH-equivalence}
It turns out that it is possible to obtain some information on the system even without directly solving Eqs.~\eqref{eq:phisp} and~\eqref{eq:psisp}.
As we are interested in the free energy rather than in the grand potential,
we perform a Legendre transform in order to have a dependence upon $\phi_{c}$ and $\phi_{b}$.
To this aim, we need to  be able to express $\kappa$ and $\eta$ in terms of these densities.

Making use of Eqs.~\eqref{eq:phibSol} and~\eqref{eq:phicSol}, and without an explicit derivation of the solutions for $\varphi$ and $\psi$, one can express the fugacities $\eta$ and $\kappa$ in terms of $\phi_b$ and $\phi_c$ as:
\begin{eqnarray}
\eta
& = & \frac{\sqrt{q(\beta)} \, \phi_c \, e^{-d\beta\epsilon_i(\phi_b+\phi_c)}}{\sqrt{d \, (1-\phi_b-\phi_c) \, (\phi_b-\phi_c)}} \, , \label{eq:FinalEta} \\
\kappa
& = & \frac{(\phi_b-\phi_c) \, e^{\beta\epsilon_i}}{q(\beta) \, \phi_b \, (1-\phi_b-\phi_c) \, e^{2d\beta\epsilon_i(\phi_b+\phi_c)}} \, . \label{eq:FinalKappa}
\end{eqnarray}
This allows us to compute the reduced free energy per site 
\begin{equation}\label{eq:LegendreTFreeEn}
\beta f = \beta \Omega + \phi_b\ln\kappa + 2\phi_c\ln\eta
\end{equation}
where it is clear that $f$ depends on both, $\phi_b$ and $\phi_c$.
However, it is more convenient to express it in terms of $\phi$ and $\bar{\ell} = \phi/\phi_c$ ({\it i.e.}, a measure of the average number of monomers per chain):
\begin{eqnarray}\label{eq:centralresult}
\beta f
& = & -d \, \beta\epsilon_i \, \phi^2 + (1-\phi)\ln(1-\phi) \nonumber\\
& & + \phi \bigg\{ \frac1{\bar{\ell}}\ln\phi - \bigg(1-\frac{2}{\bar{\ell}}\bigg)\ln\bigg(\frac{q(\beta)}{e}\bigg) \nonumber\\
& & + \bigg( 1 - \frac1{\bar{\ell}} \bigg) \beta\epsilon_i + \bigg( 1-\frac2{\bar{\ell}} \bigg) \ln\bigg(\frac{\bar{\ell}-2}{\bar{\ell}-1}\bigg) \nonumber\\
& & - \frac1{\bar{\ell}}\ln\bigg(\frac{d \, \bar{\ell} \, (\bar{\ell}-1)}e \bigg) \bigg\} \, ,
\end{eqnarray}
where $e=2.71828...$ is the Euler's number.
Eq.~\eqref{eq:centralresult} is a key result of our mean-field analysis.
Notice that, since in our model the minimum length of a chain is defined to be $=1$ bond ({\it i.e.}, 2 monomers), it must be $\bar{\ell} \geq 2$.

Now that we have a mean-field estimate of the free energy, we employ Eq.~\eqref{eq:centralresult} and compute the internal energy $U$ of the system:
\begin{equation}
\label{eq:InternalEnergy}
\frac{U}{N} = \frac{\partial \beta f}{\partial \beta} = \frac{\langle N_i \rangle}{N}(-\epsilon_i) + \frac{\langle N_a \rangle}{N}\epsilon_a \, .
\end{equation}
where
\begin{eqnarray}
\frac{\langle N_i \rangle}{N}
& = & d\phi^2 - \phi\bigg(1-\frac{1}{\bar{\ell}}\bigg) \, , \label{eq:InteractionDensity} \\
\frac{\langle N_a \rangle}{N}
& = & \frac{q(\beta)-2}{q(\beta)}\bigg(1-\frac{2}{\bar{\ell}}\bigg)\phi \, . \label{eq:AngleDensity}
\end{eqnarray}
Eqs.~\eqref{eq:InteractionDensity} and~\eqref{eq:AngleDensity} bear interesting physical interpretations.
Let us first discuss Eq.~\eqref{eq:InteractionDensity}.
Within our mean-field approach the number of interactions per lattice site $\frac{\langle N_i \rangle}{N}$ is \textit{not} guaranteed to be always non-negative.
From Eq.~\eqref{eq:InteractionDensity} in fact, we readily see this to be true only when $\phi \geq \frac{1}{d}\bigg(1-\frac{1}{\bar{\ell}}\bigg)$.
On the other hand, in the limiting case $\phi=1$, it is easy to check that the density of interactions is {\it exactly} that predicted by Eq.~\eqref{eq:InteractionDensity}, namely $d-(1-1/\bar{\ell})$.
Also, the interaction density is a decreasing function of the average chain length, as it should be since the interaction is only between non-consecutive nearest neighbour monomers.
Then we focus on Eq.~\eqref{eq:AngleDensity}, noticing that the density of angles is linear in $\phi$ and the value of the bending rigidity can only modify the corresponding proportionality constant.
From the dependence of Eq.~\eqref{eq:AngleDensity} on the number of monomers per chain $\bar{\ell}$, we see that:
(i) if $\bar{\ell} = 2$ ({\it i.e.}, all chains consist of 1 bond only) the total number of angles is $=0$, as expected;
(ii) increasing $\bar{\ell}$ leads to a larger number of angles;
(iii) the proportionality constant between the angle density and $\phi$ is less than $1$ for every value of $\beta \epsilon_a$ and $\bar{\ell}$, as it should.

Further insights can be obtained by comparing our results with past work by Flory who also derived a statistical thermodynamics theory for semiflexible chain molecules on a lattice using combinatorial arguments~\cite{Flory1956a}.
A central quantity of his theory is the mean bending degree of the chain,
\begin{equation}\label{eq:degree_bending_Flory}
g_F = \frac{2(d-1)e^{-\beta\epsilon_a}}{1+2(d-1)e^{-\beta\epsilon_a}} \, ,
\end{equation}
which results to be independent of the concentration.
The same quantity can be computed (by using Eq.~\eqref{eq:AngleDensity}) within our approach as well, {\it i.e.}
\begin{equation}\label{eq:degree_bending}
    g = \frac{\langle N_a \rangle}{N}\frac{1}{\phi_c(\bar{\ell}-2)} = \frac{(d-1)e^{-\beta\epsilon_a}}{1+(d-1)e^{-\beta\epsilon_a}} \, ,
\end{equation}
is the average monomer fraction where the polymer chain displays a turn.
As a direct consequence of Eq.~\eqref{eq:AngleDensity}, our estimate of $g$ is also independent of the concentration $\phi$, in agreement with Flory.
This means that the degree of bending is only dependent on the temperature, irrespective of whether the chain is in a melt or in a dilute solution~\cite{g-does-depend-on-phi}.

Note that a comparison between the two estimates gives $g < g_F$.
On the other hand we can also argue that $g_F$ is clearly overestimating the true value because it means that at each position along the chain, the possible directions available to make a turn are always $2(d-1)$, thereby not accounting for the long-range correlations due to self-avoidance.
By contrast, Eq.~\eqref{eq:degree_bending} implies that the number of possible directions to make a turn at each step along the chain is on average $d-1$, thus effectively accounting for possible long-range correlations induced by self-avoidance.

As both estimates Eqs.~\eqref{eq:degree_bending_Flory} and~\eqref{eq:degree_bending} are independent of the concentration as remarked, it seems not plausible that it may act as an order parameter for a phase transition.
Nevertheless, Flory postulates the existence of a phase-separated, ordered state ($g = 0$) also at $T > 0$, and assumes that the entropy of such state is 0.
He then proceeds in computing the free energy difference between the latter and a completely mixed, disordered state ($g \neq 0$), the free energy of which is computed within the same mean-field theory.
He then verifies {\it a posteriori} that there exists a critical temperature below which the ordered state is thermodinamically more stable.
The premise that the entropy of the ordered state is $0$ is crucial for the derivation of Flory, and it is ultimately this assumption that leads to the appearance of a critical temperature, corresponding to a first-order phase transition.
Later studies by Gujrati and coworkers~\cite{Gujrati1980a,Gujrati1981a,Gujrati1982a} were however able to derive an {\it exact} lower bound for the entropy of a system of self-avoiding chains on a lattice that was found to be strictly positive at any temperature $T > 0$, therefore proving that a completely ordered state cannot exist unless $T=0$.

Another interesting point is related to the free energy difference between a mixed and a phase separated state at the same temperature and, therefore, at the same $g$.
This free energy difference is calculated as
\begin{equation}\label{eq:mixing}
\Delta f(\phi,\bar{\ell}) = f(\phi,\bar{\ell})-\phi f(1,\bar{\ell}) - (1-\phi)f(0,\bar{\ell})
\end{equation}
By using Eq.~\eqref{eq:centralresult}, we get 
\begin{equation}\label{eq:floryhuggins}
    \beta\Delta f(\phi,\bar{\ell}) = d \, \beta\epsilon_i \, \phi(1-\phi) + \frac{1}{\bar{\ell}}\phi\ln\phi + (1-\phi)\ln(1-\phi) \, ,
\end{equation}
which is essentially identical to the result of the Flory-Huggins (FH) model~\cite{Huggins1942,Flory1956a,Rubinstein2003,Doi2013}.

Eq.~\eqref{eq:floryhuggins} deserves some comments.
The energetic term in the original FH model~\cite{Rubinstein2003,Doi2013} also includes a (Flory) parameter accounting for the polymer-solvent and the solvent-solvent interaction.
Within our field theory, it is not difficult to account too for the polymer-solvent interactions by
modifying Eq.~\eqref{eq:Bphimu} as follows 
\begin{widetext}
\begin{equation}\label{eq:Bphimumod}
B[\{\vec{\varphi}_{\mu}(\vec x)\}]
= \frac{\kappa e^{-2(d-1) \beta\epsilon_{\rm ms}} e^{-\beta\chi}}2 \left[ (1-e^{-\beta\epsilon_a}) \sum_{\mu=1}^d |{\vec \varphi}_{\mu}(\vec x)|^2 + e^{-\beta\epsilon_a} \left( \sum_{\mu=1}^d {\vec \varphi}_{\mu}(\vec x) \right)^{2} \right] + \sqrt{\kappa} \, \eta \, e^{-(2d-1)\beta\epsilon_{\rm ms}} e^{-\beta\chi/2} \sum_{\mu=1}^d \varphi_{\mu}^1(\vec x) \, .
\end{equation}
\end{widetext}
In Eq.~\eqref{eq:Bphimumod}, we have introduced $\epsilon_{\rm ms}$ as the interaction parameter between monomer and solvent and we have replaced $\epsilon_i$ with the combination $\chi =\beta \epsilon_i + 2\beta\epsilon_{\rm ms}$, which closely resembles the so called Flory parameter in the original FH formulation~\cite{Rubinstein2003,Doi2013}.
By repeating the exact same procedure of Sec.~\ref{sec:LatticeModel-MFApprox}
and by performing the same Legendre transform already discussed in this section, one finds 
\begin{equation}\label{eq:floryhuggins-2}
\beta\Delta f(\phi,\bar{\ell}) = d \chi \phi(1-\phi) + \frac{1}{\bar{\ell}}\phi\ln\phi + (1-\phi)\ln(1-\phi) \, ,
\end{equation}
where now the energetic term includes also an explicit polymer-solvent interaction.

Beyond their formal resemblance, it must be also stressed that Eq.~\eqref{eq:floryhuggins-2} is more general than the original FH theory, since it includes also the case when the system is polydisperse.
The monodisperse limiting case is selected when dividing by the {\it average} chain length $\bar{\ell}$ in the second term.
Finally, it is evident from Eq.~\eqref{eq:floryhuggins-2} that $\Delta f$ does {\it not} depend on $\epsilon_{a}$.
Although this may appear surprising at a first sight, it is a natural consequence of the fact that all the terms featuring the bending stiffness are linear in $\phi$, therefore they disappear when one computes the free energy variation as defined in Eq.~\eqref{eq:mixing}.

\section{Discussion and conclusions}\label{sec:Concls}
The goal of the present study was to shed some light on the challenging problem of predicting the phase behavior of a system of interacting semiflexible polymers in solution because of its important consequences on protein aggregations~\cite{Fuxreiter2021,MichaelsVendruscolo2023} as well as on polymer crystallization~\cite{Olmsted1998}.

To this aim, we made the following approximations.
First, we work with implicit solvent where both intrachain and interchain interactions display a short-range attraction mimicking the effect of the solvent.
Second, we work on a $d$-dimensional lattice where these attractive energies are taken to be equal and acting only between non-consecutive nearest-neighbours chain points, and bending rigidity is represented by an energy penalty attributed on each turn of a chain.
Hence the system of interacting semiflexible polymers is represented by a system of self-avoiding walks where each turn is penalized and each nearest-neighboring occurrence is rewarded.
Finally, we constructed a field theory representation of this system and solved it within a mean-field approximation.
Specifically, we have derived a mean-field solution of the grand potential $\Omega(\kappa, \eta, \epsilon_a, \epsilon_i)$ of the system (Eq.~\eqref{eq:completefreeene}) obtained by making use of a field-theoretical representation based on the polymer-magnet analogy ($\mathcal{O}(n \! \to \! 0)$-model~\cite{deGennes1972,deGennes1979,Doniach1996}).

By solving the saddle point equations~\eqref{eq:phisp} and~\eqref{eq:psisp}, we have deduced the
bond (Eq.~\eqref{eq:phibSol}),
chain (Eq.~\eqref{eq:phicSol})
and, hence, monomer (Eq.~\eqref{eq:MonomerPhi})
density as a function of the parameters of the model.
A discontinuity appears only (Secs.~\ref{sec:Entropy1Stiffness0Energy1} and~\ref{sec:Entropy1Stiffness1Energy1})
for non-zero attractive interaction between non-consecutive nearest-neighbouring monomers where there is an abrupt shift in the total monomer density $\phi$ (taken here as the order parameter) from a gas phase ($\phi \simeq 0$) to a liquid-like phase ($\phi \simeq 1$).
Notably, Grand Canonical Monte Carlo simulations of the lattice model were found in very good agreement with (and, hence, confirm) the mean-field results.
Notice that a similar gas-liquid transition was observed also in previous MC simulations of multi-chain systems~\cite{Sheng1996,Ivanov2003} and in experiments~\cite{Olmsted1998}.
Last but not least, our theory predicts (Sec.~\ref{sec:FH-equivalence})
that the free energy variation upon mixing has a Flory-Huggins-like form, yet our result is slightly more general as it accounts for the situation where the system remains polydisperse.
Otherwise, since each contribution to the free energy depending on the bending stiffness is linear in $\phi$, the free energy variation upon mixing is independent of the bending stiffness.

In principle, different transitions might be expected~\cite{Ivanov2003}.
First, a gas-liquid transition from a low density to a high density phase.
Second, a isotropic-nematic transition from a randomly oriented isotropic phase to a phase where the stiffness induces an overall tendency to align along a common director. 
Finally, a coexistence of these two might also be present as the isotropic-nematic transition can be located either on the high-density (liquid) side of the gas-liquid transition for small stiffness, or on the low-density region of the low-density (gas) side for sufficiently stiff chains.
Triple points then might also be present.

Quite surprisingly, the bending rigidity plays no role in our theory, its effect being to renormalize the bond and chain fugacity in apparent contradiction with numerical simulations~\cite{Ivanov2003} predicting instead the liquid phase to be nematic.
The origin of this discrepancy is likely to be ascribed to the fact that within our approach chains are polydisperse, and polydispersity is known to destabilize the isotropic-nematic transition~\cite{Woolston2015}.

Our approach extends previous mean-field analysis for a single chain~\cite{Doniach1996} to multi-chain systems by accounting, in an intuitive and transparent manner, for all the fundamental ingredients of polymer solutions (chain connectivity, bending stiffness, monomer-monomer interactions, role of dilution).
While few mean-field theories do exist in the literature for multi-chain systems~\cite{Flory1956a,desCloizeaux1975,GujratiMagneticAnalog1981,FreedMultiChain1985}, none of them considered the general approach proposed here.
Overall, we trust that our theory will provide a guidance toward more dedicated approaches dealing with specific cases.
For instance, a recent study~\cite{Du2022} discussed the possibility of tailoring the conditions for observing the folding of a double helix from the self-folding of a single semiflexible polymer.
Providing a guidance for navigating in the large parameter space that is usually required for such investigations would prove an invaluable tool.

{\it Acknowledgements} --
A.G. acknowledges the MIUR PRIN-COFIN2017 {\it Soft Adaptive Networks} grant 2022JWAF7Y.
A.G. and A.R. acknowledge networking support by the COST Action CA17139 (EUTOPIA).

\appendix

\section{Derivation of Eq.~\eqref{eq:geometrical_partition_function}}\label{App:Exp-BetaH-expansion}
In order to demonstrate Eq.~\eqref{eq:geometrical_partition_function}, we start by characterizing first the consequences of the geometrical average Eq.~\eqref{eq:geometrical_average} and, in particular, of the quadratic form (Eq.~\eqref{eq:moment_theorem}) of the corresponding moment-generating function in the limit $n \to 0$.
For instance, it is easy to see that the $m$-th moment of the spin component $S^i$,
\begin{equation}
\lim_{n \to 0} \langle (S^i)^m \rangle_{\Omega} = \lim_{n \to 0} \frac{\partial^m}{\partial (\varphi^i)^m} \langle e^{\vec{S}\cdot\vec{\varphi}}\rangle_{\Omega}\bigg|_{\vec{\varphi} = 0} \, ,
\end{equation}
is different from $0$ only if $m = 0$ or $m = 2$.
Similarly, the moment containing different components
\begin{equation}
\lim_{n \to 0}\langle S^i S^j \rangle_\Omega = \lim_{n \to 0} \frac{\partial}{\partial\varphi^i}\frac{\partial}{\partial\varphi^j}\langle e^{\vec{S}\cdot\vec{\varphi}}\rangle_{\Omega}\bigg|_{\vec{\varphi} = 0}
\end{equation}
gives $\delta_{ij}$.
To summarize, given the previous two equations it is not difficult to see that {\it in the limit $n \to 0$}, the geometrical average Eq.~\eqref{eq:geometrical_average} has the following properties:
\begin{eqnarray}
\langle 1 \rangle_\Omega & = & 1 \, , \label{eq:GeometricTraceProperties-1} \\
\langle S^i \rangle_\Omega & = & 0 \, , \label{eq:GeometricTraceProperties-2} \\
\langle S^i S^j \rangle_\Omega & = & \delta_{ij} \, , \label{eq:GeometricTraceProperties-3} \\
\langle S^{i_1} S^{i_2} \, ... \,  S^{i_k} \rangle_\Omega & = & 0 \, , \, \, \, \text{if $k \geq 3$} \, . \label{eq:GeometricTraceProperties-4}
\end{eqnarray}

We focus now on Eq.~\eqref{eq:geometrical_partition_function}.
By using the expression for the lattice Hamiltonian Eq.~\eqref{eq:hamiltonian-pbc} we have:
\begin{eqnarray}\label{eq:TruncatedExpansion}
\lim_{n \to 0} \langle e^{-\beta{\mathcal H}} \rangle_{\Omega}
& = & \lim_{n \to 0} \bigg\langle \prod_{\vec x} \prod_{\mu = 1}^d \bigg( 1 + J{\vec S}({\vec x}) \cdot {\vec S}({\vec x} + {\hat e}_{\mu}) \nonumber\\
& & + \frac{J^2}2 ({\vec S}({\vec x}) \cdot {\vec S}({\vec x} + {\hat e}_{\mu}) )^2 \bigg) \bigg\rangle_{\Omega} \, .
\end{eqnarray}
It is, in fact, sufficient to stop at the second term in the expansion
because of the property Eq.~\eqref{eq:GeometricTraceProperties-4}, or every term in which a spin at a given site appears more than twice gives contribution $=0$.

\begin{figure}[h]
\includegraphics[width=0.35\textwidth]{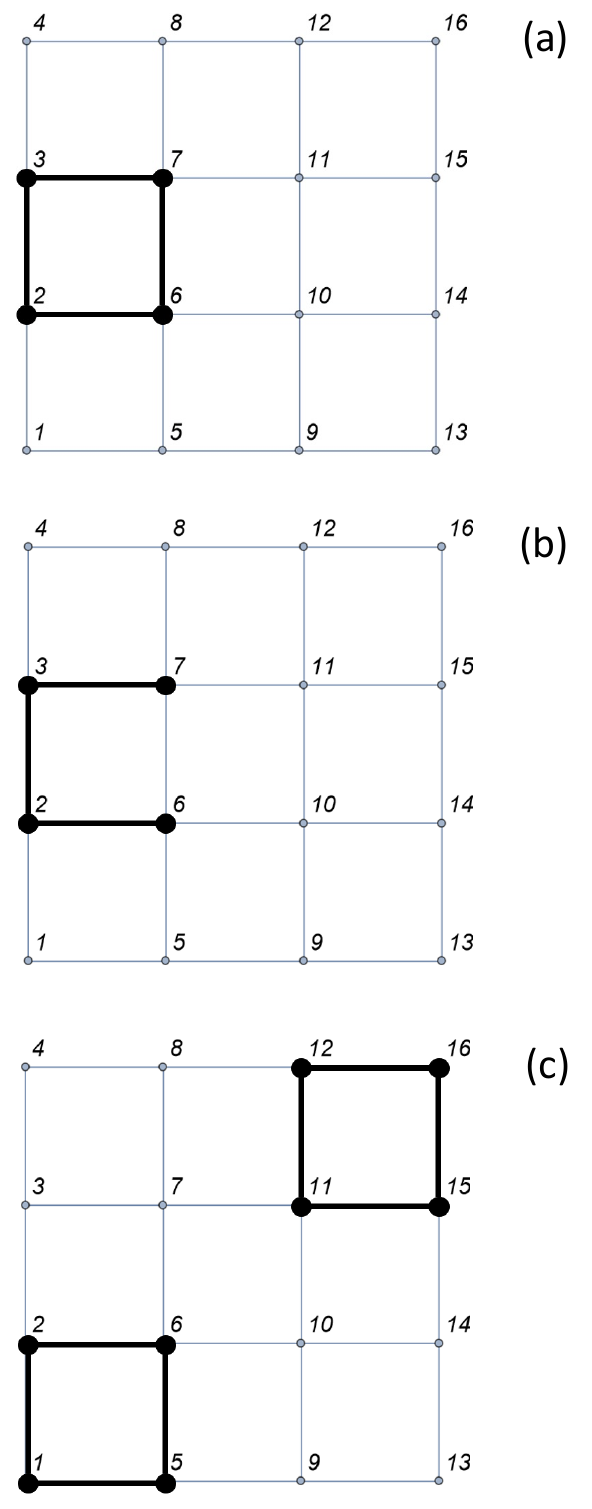}
\caption{
Examples of some possible configurations on the 2$d$ square lattice.
(a) A single closed loop.
(b) An open chain.
(c) A configuration made of two disjointed closed loops.
}
\label{fig:examples}
\end{figure}

Interestingly, we can assign the following meaning to each of the terms in Eq.~\eqref{eq:TruncatedExpansion}:
\begin{itemize}
\item
The term $1$ corresponds to an empty site;
\item
The term $\vec{S}(\vec{x})\cdot\vec{S}(\vec{x}+\hat{e}_{\mu})$ corresponds to a bond connecting sites $\vec{x}$ and $\vec{x} + \hat{e}_\mu$;
\item
The term $(\vec{S}(\vec{x})\cdot\vec{S}(\vec{x}+\hat{e}_{\mu}))^2$ correspond to a two-step closed loop between sites $\vec{x}$ and $\vec{x} + \hat{e}_\mu$.
\end{itemize}
Let us consider the particular bond configuration depicted in panel (a) of Fig.~\ref{fig:examples}.
The corresponding term is
\begin{widetext}
\begin{eqnarray}
& & J^4 \langle (\vec{S}(\vec{x}_2)\cdot\vec{S}(\vec{x}_6)) \, (\vec{S}(\vec{x}_6)\cdot\vec{S}(\vec{x}_7)) \, (\vec{S}(\vec{x}_2)\cdot\vec{S}(\vec{x}_3)) \, (\vec{S}(\vec{x}_3)\cdot\vec{S}(\vec{x}_7)) \rangle_\Omega \nonumber\\
& = & J^4 \sum_{i_1, i_2, i_3, i_4} \langle S^{i_1}(\vec{x}_2)S^{i_1}(\vec{x}_6) \, S^{i_2}(\vec{x}_6)S^{i_2}(\vec{x}_7) \, S^{i_3}(\vec{x}_2) S^{i_3}(\vec{x}_3) \, S^{i_4}(\vec{x}_3)S^{i_4}(\vec{x}_7) \rangle_{\Omega}
\end{eqnarray}
\end{widetext}
that, since spins on different sites are independent under the geometrical average $\langle \cdot \rangle_{\Omega}$, can be factorized as
\begin{widetext}
\begin{equation}\label{eq:SAL_expansion}
J^4\sum_{i_1, i_2, i_3, i_4} \langle S^{i_1}(\vec{x}_2)S^{i_3}(\vec{x}_2)\rangle_\Omega \, \langle S^{i_1}(\vec{x}_6)S^{i_2}(\vec{x}_6)\rangle_\Omega \, \langle S^{i_3}(\vec{x}_3)S^{i_4}(\vec{x}_3)\rangle_\Omega \, \langle S^{i_2}(\vec{x}_7)S^{i_4}(\vec{x}_7) \rangle_\Omega \, .
\end{equation}
\end{widetext}
It is now easy to realize that, in order for this term to be non-zero, the only possibility is to take $i_1 = i_2 = i_3 = i_4 = i$, leading to the result:
\begin{equation}
J^4\sum_{i = 1}^{n}1 = n J^4 \, .
\end{equation}
It is not difficult to extend this result and conclude that {\it every possible} self-avoiding loop of $k$ steps will appear in the expansion with a weight $n J^{k}$.
Let us now consider an {\it open} chain as, for instance, the one in panel (b) of Fig.~\ref{fig:examples} corresponding to the term:
\begin{eqnarray}
& & J^3 \langle (\vec{S}(\vec{x}_2)\cdot\vec{S}(\vec{x}_6)) \, (\vec{S}(\vec{x}_2)\cdot\vec{S}(\vec{x}_3)) \, (\vec{S}(\vec{x}_3)\cdot\vec{S}(\vec{x}_7)) \rangle_\Omega \nonumber\\
& = & J^3 \sum_{i_1, i_2, i_3} \langle S^{i_1}(\vec{x}_2)S^{i_1}(\vec{x}_6) \, S^{i_2}(\vec{x}_2)S^{i_2}(\vec{x}_3) \, S^{i_3}(\vec{x}_3)S^{i_3}(\vec{x}_7) \rangle_{\Omega} \nonumber\\
\end{eqnarray}
Again, we can factorize
\begin{widetext}
\begin{equation}
J^3\sum_{i_1, i_2, i_3} \langle S^{i_1}(\vec{x}_2)S^{i_2}(\vec{x}_2)\rangle_\Omega \, \langle S^{i_2}(\vec{x}_3)S^{i_3}(\vec{x}_3)\rangle_\Omega \, \langle S^{i_1}(\vec{x}_6)\rangle_\Omega \, \langle S^{i_3}(\vec{x}_7)\rangle_\Omega \, .
\end{equation}
\end{widetext}
Notice that, since the spins in positions $\vec{x}_6$ and $\vec{x}_7$ appear only once, because of the trace properties~\eqref{eq:GeometricTraceProperties-1}-\eqref{eq:GeometricTraceProperties-4} the weight of this configuration is $=0$. 
We conclude then, that a single-chain configuration has non-zero weight {\it if and only if} each spin ({\it i.e.}, each lattice site) appears exactly twice or does not appear at all,
in other words if the configuration corresponds to a self-avoiding {\it closed loop}.
Finally, let us consider the last scenario illustrated in panel (c) of Fig.~\ref{fig:examples}, namely two {\it disjointed} loops.
The corresponding contribution to the partition function has a similar form as of Eq.~\eqref{eq:SAL_expansion}, with $8$ occupied lattice sites instead of $4$ and where again each lattice sites appears exactly twice.
One can easily check that, of the $8$ component indices, only the indices of spins belonging to the same connected part of the configuration need to be equal in order for the term to give a non-zero contribution.
In turns, this leads to the following contribution to the partition function:
\begin{equation}
J^8 \sum_{i,j = 1}^n 1 = n^2 J^8 \, .
\end{equation}
More in general, configurations with $k$ disconnected loops have a weight proportional to $n^{k}$.
This concludes the proof of Eq.~\eqref{eq:geometrical_partition_function}.

Let us now see briefly why the definitions~\eqref{eq:IntroducingTraceProperties-1}-\eqref{eq:IntroducingTraceProperties-4} lead directly to the enumeration of Hamiltonian paths.
The crucial point is that $\langle 1 \rangle_0 = 0$.
Since spins on different sites are independent under the trace operation, the only non-zero terms in Eq.~\eqref{eq:TruncatedExpansion} are those where {\it every} spin appears exactly {\it twice}.
Based on the previous discussion, it is easy to see that such terms correspond to Hamiltonian closed paths.
Again, these terms will have a weight proportional to $n$.

To conclude, we discuss briefly the trick to count multiple open chains instead of single closed loops.
The basic idea is to introduce an external magnetic field in an arbitrary direction in the $O(n)$-model Hamiltonian.
Let us denote by $H$ the external magnetic field along the direction $1$.
The expansion of $\langle e^{-\beta\mathcal{H}}\rangle_\Omega$ can be truncated at
\begin{equation}
\lim_{n \to 0} \bigg\langle \prod_{\vec{x}}\bigg(1+H S^1(\vec{x})\bigg)\prod_{\mu = 1}^{d}\bigg(1 + J \vec{S}(\vec{x})\cdot\vec{S}(\vec{x}+\hat{e}_{\mu})\bigg) \bigg\rangle_\Omega \, .
\end{equation}
The term $H S^1({\vec x})$ now corresponds to the presence of a chain end located at site ${\vec x}$.
Thus, for instance, the configuration of Fig.~\ref{fig:examples}(b) would be described by the term
\begin{widetext}
\begin{equation}
H^2 J^3 \langle S^1(\vec{x}_6) \, ({\vec S}(\vec{x}_6) \cdot {\vec S}(\vec{x}_2)) \, ({\vec S}({\vec x}_2) \cdot {\vec S}({\vec x}_3)) \, ({\vec S}({\vec x}_3)\cdot {\vec S}({\vec x}_7)) \, S^1({\vec x}_7) \rangle_\Omega \, .
\end{equation}
\end{widetext}
Notice that now, with the introduction of the external field, each spin appears exactly twice and, therefore, the weight of the configuration is non-zero.
By proceeding with the factorization we get
\begin{widetext}
\begin{equation}
H^2 J^3 \sum_{i_1, i_2, i_3} \langle S^1({\vec x}_6) S^{i_1}({\vec x}_6) \rangle_\Omega \, \langle S^{i_1}({\vec x}_2)S^{i_2}({\vec x}_2) \rangle_\Omega \, \langle S^{i_2}({\vec x}_3)S^{i_3}({\vec x}_3)\rangle_\Omega \, \langle S^{i_3}({\vec x}_7)S^{1}({\vec x}_7) \rangle_\Omega \, .
\end{equation}
\end{widetext}
The fact that the component $1$ appears now explicitly determines a crucial difference with respect to the previous cases: in order for this term to be non-zero, all the component indices must be equal to $1$, {\it i.e.} the direction of the external magnetic field.
Therefore, only one term in the summation survives and the weight of the configuration is simply $H^2 J^3$, {\it i.e.} it is of order $n^{0}$.
It can be easily checked with the same type of calculations that the weight of a configurations with $p$ chains and $k$ total bonds is $H^{2p}J^{k}$. 
Once again, the weight of configurations consisting of closed loops remains non-zero and of order $n$.
Therefore, in the limit $n \to 0$, we account for only those configurations with no contribution from closed loops.

\section{Solutions of Eq.~\eqref{eq:phisp-Entropy1Stiffness0Energy0} for $\eta>0$}\label{App:SolutionCubic}
By noticing that the denominator on the r.h.s. of the saddle-point equation~\eqref{eq:phisp-Entropy1Stiffness0Energy0} is equal to the argument of the logarithm in the grand potential~\eqref{eq:completefreeene} and, therefore, it must be strictly positive, we may rearrange the~\eqref{eq:phisp-Entropy1Stiffness0Energy0} in the cubic form:
\begin{equation}\label{eq:cubic_sp}
p(\varphi) \equiv
\frac{d^2 \kappa}4 \varphi^3 + \frac{d \sqrt{\kappa} \, \eta}2 \varphi^2 + \bigg(\frac{1}{2}-d\kappa\bigg)\varphi - \sqrt{\kappa} \eta = 0 \, .
\end{equation}
\begin{figure}[h]
\includegraphics[width=0.45\textwidth]{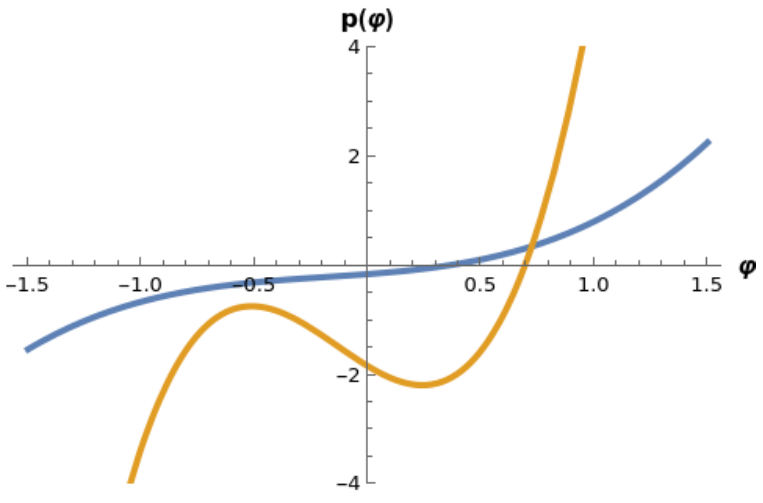}
\caption{
Two possible scenarios for the function $p(\varphi)$ in Eq.~\eqref{eq:cubic_sp}.
In the first scenario (blue line) $p(\varphi)$ is monotonically increasing, whereas in the second one (orange line) a local maximum and a local minimum appear.
In both situations, the curve intersects the positive $\varphi$-semiaxis once and only once. 
}
\label{fig:cubic}
\end{figure}

Then we notice that the coefficient of the cubic term is positive, so $p(\varphi \to -\infty) \to -\infty$ and $p(\varphi \to +\infty) \to +\infty$, while $p(0) \leq 0$.
In order to gain some insight on the possible solutions of~\eqref{eq:cubic_sp}, we study the first derivative of $p(\varphi)$ with respect to $\varphi$,
\begin{equation}\label{eq:cubic_sp_firstderivative}
p'(\varphi) = \frac{3 d^2 \kappa}{4}{\varphi^2} + d \sqrt{\kappa} \, \eta \, \varphi + \bigg(\frac12 - d\kappa\bigg) \, .
\end{equation}
By setting $p'(\varphi) = 0$, only two scenarios are possible (see Fig.~\ref{fig:cubic}):
\begin{itemize}
\item
If $\eta < \sqrt{\frac32}$ and $\kappa < \frac1d \bigg( \frac12 - \frac{\eta^2}3 \bigg)$ then $p(\varphi)$ is a monotonically increasing function.
Since $p(0) \leq 0$, Eq.~\eqref{eq:cubic_sp} has only one real solution, $>0$ for any $\kappa>0$ and $=0$ for $\kappa=0$.
\item
For all other values of $\kappa$ and $\eta$, $p'(\varphi) = 0$ has two solutions,
\begin{equation}
\varphi_{\pm} = -d\sqrt{\kappa}\eta \bigg[ 1 \pm \sqrt{ 1 + \bigg( d\kappa - \frac12 \bigg) \frac3{\eta^2} } \bigg] \, ,
\end{equation}
corresponding to a local maximum ($\varphi_+$) and a local minimum ($\varphi_-$) for $p(\varphi)$.
Since the local maximum is for $\varphi_+ < 0$, and always remembering that $p(0) \leq 0$, also in this case Eq.~\eqref{eq:cubic_sp} has only one real positive solution.
\end{itemize}
%

\section{Numerical solutions of Eqs.~\eqref{eq:phisp-Entropy1Stiffness0Energy1} and~\eqref{eq:psisp-Entropy1Stiffness0Energy1}}\label{App:ShowPhaseTransition}
It is not difficult to see that Eqs.~\eqref{eq:phisp-Entropy1Stiffness0Energy1} and~\eqref{eq:psisp-Entropy1Stiffness0Energy1} can be recast as the following two equations, both expressing $\psi$ as a function of $\varphi$ only:
\begin{widetext}
\begin{eqnarray}
\frac{\psi}{2d\sqrt{\beta\epsilon_i}}
& = & \frac{d\varphi^2}4 \, \frac{\varphi + \frac{2\eta}{d\sqrt{\kappa}}e^{\beta\epsilon_i/2}}{\varphi + \frac{\eta}{d\sqrt{\kappa}}e^{\beta\epsilon_i/2}} \, , \label{eq:ShowPhaseTrans1} \\
\frac{\psi}{2d\sqrt{\beta\epsilon_i}}
& = & \frac1{2d \, \beta\epsilon_i} \, \log\left( \frac{{\varphi}/2}{-d^2 \kappa \, e^{-\beta\epsilon_i} \, \varphi^3/4 - d\sqrt{\kappa} \, \eta \, e^{-\beta\epsilon_i/2} \, \varphi^2/2 + d\kappa \, e^{-\beta\epsilon_i} \varphi + \sqrt{\kappa} \, \eta \, e^{-\beta\epsilon_i/2}} \right) \, . \label{eq:ShowPhaseTrans2}
\end{eqnarray}
\end{widetext}
In fact Eq.~\eqref{eq:ShowPhaseTrans1} follows from the ratio of the original two, while Eq.~\eqref{eq:ShowPhaseTrans2} can be obtained from Eq.~\eqref{eq:phisp-Entropy1Stiffness0Energy1} by extracting $\psi$ as a function of $\varphi$.
Then, it is an elementary exercise to solve numerically the system of these two equations for given values of $d$, $\eta$, $\kappa$ and $\epsilon_i$.

%
\begin{figure*}
\includegraphics[width=1.00\textwidth]{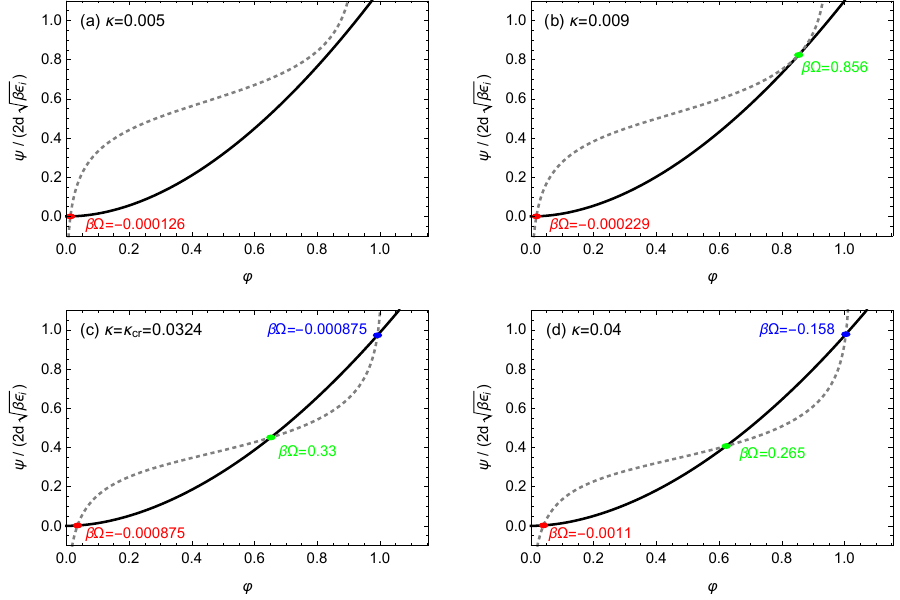}
\caption{
Numerical solutions (large dots) for the system of Eqs.~\eqref{eq:ShowPhaseTrans1} (solid lines) and~\eqref{eq:ShowPhaseTrans2} (dashed lines) in the physically admissible interval $\varphi\in[0, \sqrt{4/d}]$ (see Eq.~\eqref{eq:phibSol}) for $d=3$, $\eta=0.15$ and $\epsilon_i / k_BT=1$ (see Sec.~\ref{sec:Entropy1Stiffness0Energy1} and Fig.~\ref{fig:bond_ei+chain_ei}) and four representative values of the bond fugacity $\kappa$ (see legends, notice the ``critical'' value $\kappa_{\rm cr}$ of panel (c)).
The corresponding values for the grand potential per lattice site $\beta\Omega$ (Eq.~\eqref{eq:completefreeene}) are also shown.
}
\label{fig:ShowPhaseTransition}
\end{figure*}
%

As an example -- and without lack of generality -- we consider again (Sec.~\ref{sec:Entropy1Stiffness0Energy1}) the three-dimensional case study with $\epsilon_i/\kappa_BT=1$ and chain fugacity $\eta=0.15$ (red lines in Fig.~\ref{fig:bond_ei+chain_ei}).
Then, based on the value of the the bond fugacity $\kappa$, four scenarios are possible (see Fig.~\ref{fig:ShowPhaseTransition}).
For low $\kappa$ (panel (a)) only one solution exists, corresponding to the gas-like phase (namely, $\phi_b \simeq \phi_c \ll 1$).
At some intermediate $\kappa$ (panel (b)), another solution appears yet the most stable one ({\it i.e.}, with the lowest grand potential $\beta\Omega$ (Eq.~\eqref{eq:completefreeene})) remains the gas-like phase.
Finally (panel (d)), for $\kappa$ larger than some ``critical'' $\kappa_{\rm cr}$ (panel (c)) the most stable solution corresponds to the liquid-like one (namely, $\phi=\phi_b+\phi_c \simeq 1$).
For other chain fugacities (as well as in the case of polymers with non-zero bending stiffness, Sec.~\ref{sec:Entropy1Stiffness1Energy1}) the picture remains essentially the same, thus justifying Figs.~\ref{fig:bond_ei+chain_ei} and~\ref{fig:bond_ebi+chain_ebi}.

\section*{Data availability}
The data that support the findings of this study are available from the corresponding author upon reasonable request.


\begin{thebibliography}{10}
\bibitem{WhyEtaSquareNote}
The reason to adopt ``$\eta^2$'' for the chain fugacity is because each linear
  chains possesses two free ends, each of which may act as a polymerization
  unit.
\bibitem{g-does-depend-on-phi}
A preliminary analysis (data not shown) of polymer conformations generated by
  our GCMC algorithm seems to suggest that, on the contrary, $g$ depends on
  $\phi$ and that, therefore, both Eq.~\eqref{eq:degree_bending_Flory}
  and~\eqref{eq:degree_bending} are qualitatively wrong. At the same time, a
  more systematic comparison between theory and simulations would require using
  $\phi$ and $\bar{\ell}$ as tunable parameters and, hence, a different
  numerical scheme. This goes beyond the present scopes and will be the subject
  of future research work.
\end{thebibliography}



\end{document}